\documentclass[journal]{IEEEtran}
\usepackage{amsmath,amsfonts}
\usepackage{amssymb}
\usepackage{algorithmicx,algorithm}
\usepackage{array}
\usepackage{textcomp}
\usepackage{stfloats}
\usepackage{url}
\usepackage{verbatim}
\usepackage{graphicx}
\usepackage{cite}
\usepackage{amsthm}
\usepackage{theorem}
\usepackage{subfigure}
\usepackage{color}
\usepackage{multirow}

\begin{document}

\title{Grouping Method for mmWave Massive MIMO System: Exploitation of Angular Multiplexing Gain}

\author{Peng Jiang, Pengcheng Zhu, Jiamin Li and Dongming Wang
\thanks{This paper was produced by the IEEE Publication Technology Group.}
\thanks{ }}

\markboth{Journal of \LaTeX\ Class Files,~Vol.~14, No.~8, August~2021}%
{Shell \MakeLowercase{\textit{et al.}}: A Sample Article Using IEEEtran.cls for IEEE Journals}


\maketitle

\begin{abstract}
A future millimeter-wave (mmWave) massive multiple-input and multiple-output (MIMO) system may serve hundreds or thousands of users at the same time; thus, research on multiple access technology is particularly important.
Moreover, due to the short-wavelength nature of a mmWave, large-scale arrays are easier to implement than microwaves, while their directivity and sparseness make the physical beamforming effect of precoding more prominent.
In consideration of the mmWave angle division multiple access (ADMA) system based on precoding, this paper investigates the influence of the angle distribution on system performance, which is denoted as the angular multiplexing gain.
Furthermore, inspired by the above research, we transform the ADMA user grouping problem to maximize the system sum-rate into the inter-user angular spacing equalization problem.
Then, the form of the optimal solution for the approximate problem is derived, and the corresponding grouping algorithm is proposed.
The simulation results demonstrate that the proposed algorithm performs better than the comparison methods.
Finally, a complexity analysis also shows that the proposed algorithm has extremely low complexity. 

\end{abstract}

\begin{IEEEkeywords}
Massive MIMO, user scheduling, ADMA grouping, millimeter-wave communication, digital beamforming.
\end{IEEEkeywords}

\section{Introduction}\label{Sec1}
\IEEEPARstart{M}{illimeter-wave} (mmWave) massive multiple-input and multiple-output (MIMO) communication is expected to be the key technology for future mobile communication systems \cite{Chen2020MWC,Giordani2020MCOM,You2020SCIS,Matthaiou2021MCOM}.
There are plenty of available spectra in the mmWave band from 30 GHz to 300 GHz, which could be used for mobile communication.
Since a mmWave has a shorter wavelength than a microwave, it gains a higher path loss and attenuation, making deploying mmWaves difficult \cite{Heath2016JSTSP, Yu2016JSTSP}.
Fortunately, due to the short wavelength of mmWave, many mmWave antennas can be accommodated in the same space.
Hence, large-scale mmWave antenna arrays are easier to deploy, which goes well with massive MIMO technology \cite{Hong2021JMW, Heath2016JSTSP, Sohrabi2021JSAC}.
Moreover, the array gain brought by large-scale arrays can compensate for the high loss of mmWaves at a certain extent.

The precoding or beamforming design is the key to obtaining the multiantenna gain for massive MIMO systems.
Through the reasonable design of precoding, we can improve the signal strength of users and reduce the interference between users to obtain better system performance.
Moreover, massive MIMO systems with precoding can transmit independent data streams to different users at the same time and using the same frequency resources, which makes precoding the core of space division multiple access (SDMA) in centimeter-wave systems \cite{Fuschini2019TAP, Gomez2022TWC, Yoo2006JSAC} or angle division multiple access (ADMA) in mmWave systems with highly angular dependent channels \cite{Lin2017JSAC,Liu2019TCS,Wang2019JSTSP}.

A large and growing body of literature has investigated the hybrid precoding system of mmWave systems \cite{Heath2016JSTSP,Yu2016JSTSP,Hong2021JMW,Sohrabi2021JSAC,Fuschini2019TAP,Gomez2022TWC,Lin2017JSAC,Zhao2017TWC,Paul2021TMC,Koc202ACCESS}.
Due to the limitation of the current technology level, a hybrid precoding system requires a compromise between performance and implementation difficulty \cite{Heath2016JSTSP, Zhao2017TWC}. 
Although the implementation complexity is high, full-digital precoding with better performance is still an important direction of mmWave system research, and many researchers have realized the design of full-digital transmitters and achieved impressive results \cite{Yang2018TMTT, Guo2020LAWP}.
Furthermore, massive Machine Type of Communication (mMTC) \cite{Sharma2020COMST, Chen2021JSAC} requires more than one million links per square kilometer, and it is difficult for one single multiple access system to meet such a demand.
Hence, combinations of multiple access technologies should be applied, where a typical solution is to combine multiple access methods, such as combining time division multiple access (TDMA) or frequency division multiple access (FDMA) with ADMA, to group users \cite{Yoo2006JSAC, Zhao2017TWC,Lakshmi2021TCOMM, Paul2021TMC}.
Different groups of users are served with different time slots or frequencies, and users in the same group are served by ADMA at the same time and with the same frequency.

For the highly angular correlated channels in mmWave systems, the rational allocation of users can effectively improve the sum-rate of an ADMA system\cite{Zhao2017TWC}.
Previous work has focused on centimeter-wave MIMO and has used channel correlation, greedy algorithms, heuristic algorithms, and other algorithms \cite{Yoo2006JSAC, Chen2020TWC, Mauricio2021ACCESS, Lee2018TCOMM, Lakshmi2021TCOMM} to improve the sum-rate of SDMA or ADMA systems.
Although these methods can provide good performance, their algorithmic complexity is often very high and expands rapidly with an increase in the number of users. 
The user grouping problem of mmWave MIMO is considered in some innovative and practical works \cite{Paul2021TMC, Koc202ACCESS, Zhao2017TWC}.
However, most of the grouping algorithms for mmWave systems rely on hybrid precoding architectures and have performance approaching but not exceeding that of full-digital systems.
In addition, except for a few algorithms such as the greedy algorithm in \cite{Zhao2017TWC}, most of these algorithms that rely on hybrid precoding architectures cannot be applied to full-digital systems.

Nonetheless, few studies have investigated the effect of angle distribution on precoding performance in mmWave systems.
Therefore, in this paper, a more efficient grouping algorithm for the ADMA system is proposed according to the angle distribution.
The major contributions are as follows.
\begin{itemize}
\item {We first analyze the beamforming effect of linear precoding in a mmWave massive MIMO system under a uniform linear array (ULA).
The results clarify that linear precoding can concentrate the electromagnetic wave signal in a small angular range to form a narrow beam that is directed to the intended user.
In addition, the influence of the angle distribution on the system sum-rate is studied, and is later defined as the angular multiplexing gain.
}
\item{Furthermore, we transform the problem of maximizing the sum-rate of the system through reasonable user groupings into a problem of equalizing the angular spacing of users in each group by the rule of angular multiplexing gain.
Then, we prove and propose an angular spacing equalization grouping (ASEG) algorithm, which is optimal for the transformed problem.}
\item{Finally, the simulation results and the algorithm complexity analysis show that our algorithm can effectively improve the system sum-rate with an extremely low complexity while the number of users is large.}
\end{itemize}

The rest of the paper is organized as follows.
Section \ref{Sec2} describes the mmWave massive MIMO system.
Section \ref{Sec3} analyzes the beamforming effect of linear precoding and the effect of the angular distribution on the system rate.
Section \ref{Sec4} states the optimization problem and transforms it into an approximation problem.
Section \ref{Sec5} proposes the grouping algorithm.
Then, the angular multiplexing gain and proposed algorithm are simulated and analyzed in Section \ref{Sec6}.
Finally, Section \ref{Sec7} concludes this paper.

{\bf{Notations}:}
$(\cdot)^*$, $(\cdot)^\mathrm{T}$, $(\cdot)^\mathrm{H}$ and $(\cdot)^{\dag}$ refer to the conjugate, transpose, Hermitian transpose, and Moore-Penrose pseudoinverse, respectively.
$\|\cdot\|_F$ denotes the 2-norm for vectors and the Frobenius norm for matrices.
$\mathcal{CN}(0,\sigma^2)$ indicates a complex Gaussian distribution with a zero mean and a variance of $\sigma^2$.
$\mathbf{I}_N$ is an $N \times N$ identity matrix.

\section{System Model and Channel Characteristics}\label{Sec2}
\subsection{System Model}
In this paper, we consider a mmWave full-digital time division duplex (TDD) massive MIMO system in a single cell, where $K$ single-antenna users are served by a base station equipped with an $N$-antenna ULA.
Considering the actual three-sector deployment of the antenna array, we focus on the users who are located in the range within the front $2\pi/3$.

To satisfy the connection density requirement of mMTC, we need hybrid multiple access technologies.
Therefore, in this article, we chose ADMA to cooperate with ideal TDMA or FDMA to serve massive numbers of users at the same time.

{\emph{Remark 1:}} ADMA is one kind of SDMA.
In the centimeter band, we usually refer to the multiple access method based on precoding as SDMA.
However, in the mmWave band, we call it ADMA because of the high degree of correlation between precoding and the angle, which is reflected in \cite{Lin2017JSAC} and the analysis that follows in this article.

\subsection{Channel Characteristics}
The downlink channel of user $k$ to an $N$-antenna ULA can be written in the widely used Saleh-Valenzuela mmWave channel model \cite{Heath2016JSTSP,Zhao2017TWC,Lin2017JSAC} as
\begin{equation}\label{hDLk}
    \mathbf{h}_{k}=\beta^{(0)}_k {\mathbf{a}^{(0)}_k}+\sum^L_{l=1}\beta^{(l)}_k {\mathbf{a}^{(l)}_k},
\end{equation}
where $\beta^{(l)}_k$ is the complex gain of the $l$-path of user $k$, which consists of the path loss and shadow fading, and $\mathbf{a}^{(l)}_k\in\mathbb{C}^{1\times N}$ are steering vectors.
$l=0$ represents the line-of-sight (LOS) component of the user, while $l \geq 1$ represents the nonline-of-sight (NLOS) components.
The steering vectors $\mathbf{a}^{(l)}_k$ can be modeled as
\begin{equation}\label{aphak}
{\mathbf{a}^{(l)}_k}=
\left[
  \begin{array}{c}
    e^{j \cdot 2 \cdot \pi \cdot 0 \cdot \xi^{(l)} _{k}} \\
    e^{j \cdot 2 \cdot \pi \cdot 1 \cdot \xi^{(l)} _{k}} \\
    \vdots\\
    e^{j \cdot 2 \cdot \pi \cdot (N-1) \cdot \xi^{(l)} _{k}} \\
  \end{array}
\right]^{\mathrm{T}},
\end{equation}
where $\xi^{(l)} _{k}=\frac{d}{\lambda}\cos \theta^{(l)}_k$. 
$\theta^{(l)}_k$ is the physical direction of arrival (DOA) of the $l$-path of user $k$, $\lambda$ is the wavelength of the transmitted signal and $d$ is the antenna distance, which is generally set as $d=\frac{\lambda}{2}$ in a large-scale multiantenna array.

Due to the directionality of mmWaves, the NLOS components are usually much lower than the LOS elements \cite{Heath2016JSTSP, Zhang2020SCIS}.
For simplicity, researchers generally only consider the LOS components \cite{Zhao2017TWC,Zhang2017JSAC,Rupasinghe2019JSTSP,Pang2021SCIC,Sohrabi2021JSAC} in their analysis, so we can ignore the NLOS components and superscript $l$.
These NLOS components will be reconsidered in simulations.
\subsection{Downlink Data Transmission}
The downlink channel matrix of the cell can be expressed as
\begin{equation}
\mathbf{H}=\left[ \mathbf{h}^\mathrm{T}_1,\mathbf{h}^\mathrm{T}_2,\ldots,\mathbf{h}^\mathrm{T}_K\right],
\end{equation}
For massive MIMO systems, precoding is the key to multiantenna signal processing.
The precoding matrix is set as $\mathbf{P}=\left[\mathbf{p}_1,\mathbf{p}_2,...,\mathbf{p}_K\right]\in\mathbb{C}^{N\times K}$; then, the received signal vector of all users $\mathbf{Y}\in\mathbb{C}^{K\times 1}$ is given by
\begin{equation}\label{Y}
    \mathbf{Y}=\mathbf{H}\sqrt{p}\mathbf{P}\mathbf{x}+\mathbf{n},
\end{equation}
where $\mathbf{x}\in\mathbb{C}^{K\times 1}$ represents the data streams of $K$ users, $\mathbf{n}\in\mathbb{C}^{K\times 1}\sim\mathcal{CN}(0,\sigma^2 \mathbf{I}_K)$ is the noise vector and $p$ is the total power.

In the multiplexing system, we divide users into $G>1$ multiple access groups.
We use the ideal TDMA or FDMA and ADMA in each group.
Setting $\mathcal{U}$ as a universal set, we have a constraint that can be explicitly expressed as
\begin{subequations}\label{Grule}
\begin{align}
    \mathcal{U}&=\{1,2,...,K\}\\
    \mathcal{U}&=\mathcal{U}_1\cup\mathcal{U}_2\cup...\cup\mathcal{U}_G \\
    \forall i&\neq k,\mathcal{U}_i\cap\mathcal{U}_k=\emptyset,
\end{align}
\end{subequations}
In this multiplexing case, the rate of the $k$-th user in group $\mathcal{U}_g$ can be written as
\begin{equation}\label{kgRate}
    R_k=\log_2\left( 1+\frac{p\|\mathbf{h}_k\mathbf{p}_k\|_F^2}
    {\sum_{i\neq k}^{i\in \mathcal{U}_g}p\|\mathbf{h}_k\mathbf{p}_i\|_F^2+\sigma^2}\right).
\end{equation}
Then, the sum-rate of group $\mathcal{U}_g$ is
\begin{equation}\label{cgRate}
    R_{\mathcal{U}_g}=\sum_{k\in \mathcal{U}_g}R_k.
\end{equation}
Assume that time or frequency resources are allocated among groups by $S_g>0$, and $\sum_{g=1}^{G}S_g=1$; hence, the sum-rate of the whole system is given by
\begin{equation}\label{cRate}
    R=\sum_{g=1}^{G}S_g R_{\mathcal{U}_g}.
\end{equation}

In general, to ensure fairness, we assume that resources are evenly distributed across groups with the same number of users in each group.

\section{Physical Beamforming Effect of Precoding and Angular Multiplexing Gain}\label{Sec3}
Linear precoding can play the role of beamforming in massive MIMO.
Due to the directionality and angle-related channel in a mmWave system, the physical beamforming effect of linear precoding is more obvious.

This section mainly analyzes the beamforming effect brought by linear precoding in mmWave MIMO systems.
The beam formed by linear precoding has a narrow-angle, which is determined by the angles of users; thus, we correlate the system performance with the angular distribution of users.

{\emph{Remark 2:}} In this article, a beam refers to electromagnetic waves whose energy is concentrated in a small angular range, beamforming refers to the process of generating beams, and beamforming effects are the effects of forming beams through beamforming processes.

\subsection{Typical Linear Precoding Methods}
There are three typical linear precoding methods and they are derived in different ways: maximum ratio transmission (MRT), zero forcing (ZF), and minimum mean-square error (MMSE) \cite{Kebede2022ACCESS,Fatema2017JSYST}.

MRT precoding aims to maximize the rate, and regardless of interference, we can obtain the MRT precoding matrix $\mathbf{P}'_{\mathrm{MRT}}$ and the normalized MRT precoding matrix $\mathbf{P}_{\mathrm{MRT}}$, namely,
\begin{equation}\label{MRT}
\begin{array}{l}
\mathbf{P}'_{\mathrm{MRT}}=\mathbf{H}^{\mathrm{H}} \\
\mathbf{P}_{\mathrm{MRT}}=\mathbf{P}'_{\mathrm{MRT}}/\|\mathbf{P}'_{\mathrm{MRT}}\|_F.
\end{array}
\end{equation}

ZF aims to completely eliminate inter-user interference regardless of the rate; i.e.,
\begin{equation}\label{ZF}
\begin{array}{l}
\mathbf{P}'_{\mathrm{ZF}}=\mathbf{H}^{\mathrm{H}}\left( \mathbf{H}\mathbf{H}^{\mathrm{H}} \right)^{-1}\\
\mathbf{P}_{\mathrm{ZF}}=\mathbf{P}'_{\mathrm{ZF}}/\|\mathbf{P}'_{\mathrm{ZF}}\|_F.
\end{array}
\end{equation}

MMSE aims to maximize the signal-to-leakage-noise ratio (SLNR), which is a concession of the signal-to-interference-noise ratio (SINR). Hence,
\begin{equation}\label{MMSE}
\begin{array}{l}
\mathbf{P}'_{\mathrm{MMSE}}=\mathbf{H}^{\mathrm{H}}\left( \mathbf{H}\mathbf{H}^{\mathrm{H}}+\sigma^2\mathbf{I}_K \right)^{-1} \\
\mathbf{P}_{\mathrm{MMSE}}=\mathbf{P}'_{\mathrm{MMSE}}/\|\mathbf{P}'_{\mathrm{MMSE}}\|_F.
\end{array}
\end{equation}
\subsection{Physical Beamforming Effect of MRT }\label{PreCod_BF}
We consider a pair of users $k$ and $j$, where $k \neq j$, which are 
served by MRT precoding.
Their channels can be written as
\begin{equation}\label{hk}
\mathbf{h}_{k}=\beta_k
\left[
  \begin{array}{c}
    e^{j \cdot \pi \cdot 0 \cdot \cos \theta_k} \\
    e^{j \cdot \pi \cdot 1 \cdot \cos \theta_k} \\
    \vdots\\
    e^{j \cdot \pi \cdot (N-1) \cdot \cos \theta_k} \\
  \end{array}
\right]^\mathrm{T}.
\end{equation}
Then, the MRT precoding vector can be expressed as
\begin{equation}\label{MRTk}
\mathbf{p}_k=\frac{\beta^*_k}{\sqrt{N}|\beta_k|}
\left[
  \begin{array}{c}
    e^{-j \cdot \pi \cdot 0 \cdot \cos \theta_k} \\
    e^{-j \cdot \pi \cdot 1 \cdot \cos \theta_k} \\
    \vdots\\
    e^{-j \cdot \pi \cdot (N-1) \cdot \cos \theta_k} \\
  \end{array}
\right].
\end{equation}
In general, the signal sent by precoding $\mathbf{p}_k$ is the signal to user $k$ but interferes with user $j$ when $j \neq k$.
Notably, the signal amplitude of user $k$ is
\begin{equation}\label{SA_k}
    \mathrm{SA}_k=\mathbf{h}_{k}\mathbf{p}_k=\sqrt{N}|\beta_k|,
\end{equation}
and the interference amplitude to user $j$ is
\begin{equation}\label{IA_k}
    \mathrm{IA}_k=\mathbf{h}_{j}\mathbf{p}_k=
    \frac{\beta_j \beta^*_k}{\sqrt{N}|\beta_k|}
    \sum_{n=0}^{N-1}e^{j \cdot \pi \cdot n \cdot (\cos \theta_j -\cos \theta_k)},
\end{equation}

If only the relative amplitude of the signal is considered, we have
\begin{equation}\label{Phi}
\Phi(j,k) = \begin{cases}
\sqrt{N},&\theta_j=\theta_k\\ 
\frac{1}{\sqrt{N}}\frac{1-e^{j \cdot \pi \cdot N \cdot (\cos \theta_j -\cos\theta_k)}}{1-e^{j \cdot \pi (\cos \theta_j -\cos \theta_k)}},&{\text{otherwise}}
\end{cases},
\end{equation}
which represents the signal that is sent to user $k$ but received by user $j$.
\eqref{Phi} is only related to $\theta_j$ and $\theta_k$.
\begin{figure}[!t]
\centering
\includegraphics[width=3.4in]{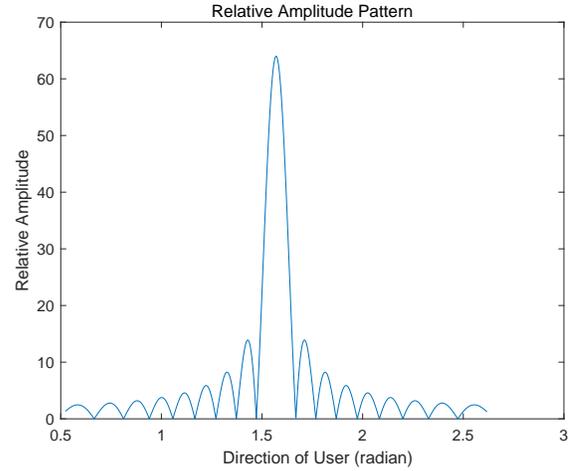}
\caption{The relationship between the signal amplitude and direction.}
\label{am_pat}
\end{figure}

We depict the relative amplitude of the signal in Fig. \ref{am_pat}.
Without loss of generality, we set $\theta_k=\frac{1}{2}\pi$ and $\theta_j$ to vary from $\frac{1}{6}\pi$ to $\frac{5}{6}\pi$ as the x-axis variable.
As shown in Fig. \ref{am_pat}, \eqref{Phi} reaches the maximum value when $\theta_j=\theta_k$ and roughly decreases when $|\theta_j -\theta_k|$ increases.
Furthermore, Fig. \ref{am_pat} reveals the beamforming effect of MRT
and the physical explanation of MRT precoding: MRT precoding sends the signal to the user in a centralized manner and reduces the leakage in other directions.
However, if other users are close to the target users, the MRT precoding beam will interfere with their signals, which must be avoided.


\subsection{Physical Beamforming Effect of ZF}
\subsubsection{ZF of a pair of users}\label{pairZF}
ZF precoding, which aims to zeroize interuser interference, is usually effective.
However, when the angular spacing of users is too small, the cost of ZF will be expensive, and the signal strength will decay sharply.
For ZF precoding, the signal pattern is more complex. 
Now, we introduce some notations to simplify the expression.
\begin{subequations}\label{Shorthand}
\begin{align}
    \phi_j & =e^{j \pi \cos\theta_j} \\
    D_{j,k} & =\frac{\phi_j}{\phi_k} \\
    t_{j,k} & =\cos\theta_j-\cos\theta_k.
\end{align}
\end{subequations}

We first calculate part of ZF; we set
\begin{equation}
\mathbf{H}_{\mathrm{z}}
=
\left[
  \begin{array}{c}
  \mathbf{h}_j\\
  \mathbf{h}_k
  \end{array}
\right].
\end{equation}
Then,
\begin{equation}\label{HHT}
\mathbf{H}_{\mathrm{z}}\mathbf{H}_{\mathrm{z}}^{\mathrm{H}}
=
\left[
  \begin{array}{cc}
  \beta_j\bar{\beta_j} N & \beta_j \bar{\beta_k} \sum^{N-1}_{n=0}D_{j,k}^n \\
  \bar{\beta_j}\beta_k \sum^{N-1}_{n=0}D_{k,j}^n & \beta_k \bar{\beta_k} N \\
  \end{array}
\right].
\end{equation}

For a second-order matrix, the inverse matrix can be obtained directly, similar to\eqref{HHT^-1}, which is at the top of the next page.

\begin{figure*}
\begin{equation}\label{HHT^-1}
\left [\mathbf{H}_{\mathrm{z}}\mathbf{H}_{\mathrm{z}}^{\mathrm{H}}\right ]^{-1}
=
\left\{ |\beta_j|^2|\beta_k|^2 \left[N^2-\frac{1-D_{j,k}^N}{1-D_{j,k}}\frac{1-D_{k,j}^{N}}{1-D_{k,j}} \right] \right\}^{-1}
\left[
  \begin{array}{cc}
  |\beta_k|^2 N & -\beta_j \beta^*_k \frac{1-D_{j,k}^N}{1-D_{j,k}} \\
  -\beta^*_j \beta_k \frac{1-D_{k,j}^N}{1-D_{k,j}} & |\beta_j|^2 N \\
  \end{array}
\right].
\end{equation}
\noindent\rule[0.25\baselineskip]{\textwidth}{0.8pt}
\end{figure*}
Then, we simplify the coefficient of \eqref{HHT^-1} as
\begin{equation}\label{INV}
\begin{aligned}
    \mathrm{INV}_{j,k} & = |\beta_j|^2|\beta_k|^2 \left[N^2-\frac{1-D_{j,k}^N}{1-D_{j,k}}\frac{1-D_{k,j}^{N}}{1-D_{k,j}}\right]\\
    &=|\beta_j|^2|\beta_k|^2 \left[N^2-\Omega(t_{j,k})\right].
\end{aligned}
\end{equation}

Considering the normalized precoding power, we have
\begin{equation}\label{norm}
\begin{aligned}
    \|\mathbf{P}'_{\mathrm{ZF}}\|_F &=\sqrt{\mathrm{Tr}(\mathbf{P}_{\mathrm{ZF}}^\mathrm{'H} \mathbf{P}'_{\mathrm{ZF}})}\\
    &=\frac{\sqrt{N}}{\sqrt{\mathrm{INV}_{j,k}}}\sqrt{|\beta_j|^2+|\beta_k|^2}.
\end{aligned}
\end{equation}
Subsequently, ZF could be rewritten as \eqref{ZF_jk}, which is shown at the top of the next page.
\begin{figure*}
\begin{equation}\label{ZF_jk}
    \mathbf{P}_{ZF}=
    \frac{1}{\sqrt{2N\cdot\mathrm{INV}_{j,k}}\sqrt{|\beta_j|^2+|\beta_k|^2}} 
    \left \{ 
    |\beta_k|^2 \beta^*_j \left [ N \mathbf{a}_j^\mathrm{H} - \frac{1-D_{k,j}^N}{1-D_{k,j}} \mathbf{a}_k^{\mathrm{H}}\right ],
    |\beta_j|^2 \beta^*_k\left [ N \mathbf{a}_k^\mathrm{H} - \frac{1-D_{j,k}^N}{1-D_{j,k}} \mathbf{a}_j^{\mathrm{H}}\right ]
    \right \}.
\end{equation}
\noindent\rule[0.25\baselineskip]{\textwidth}{0.8pt}
\end{figure*}

Finally, the precoding vector of user $j$ is
\begin{equation}\label{pair_user_zf}
\mathbf{p}_j =\frac
{|\beta_k|^2 \beta^*_j\left [ N \mathbf{a}_j^\mathrm{H} - \frac{1-D_{k,j}^N}{1-D_{k,j}} \mathbf{a}_k^{\mathrm{H}} \right ]}
{\sqrt{2N\cdot\mathrm{INV}_{j,k}}\sqrt{|\beta_j|^2+|\beta_k|^2}},
\end{equation}
and the received signal of user $j$ under ZF precoding is
\begin{equation}\label{user_j_zf}
\begin{aligned}
    \mathbf{h}_j\mathbf{p}_j & =\frac
    {\beta_j \mathbf{a}_j |\beta_k|^2 \beta^*_j\left [ N \mathbf{a}_j^\mathrm{H} - \frac{1-D_{k,j}^N}{1-D_{k,j}} \mathbf{a}_k^{\mathrm{H}}\right ]}
    {\sqrt{2N\cdot\mathrm{INV}_{j,k}}\sqrt{|\beta_j|^2+|\beta_k|^2}} \\
    &=\frac{|\beta_j||\beta_k| \sqrt{N^2-\Omega(t_{j,k})}}
    {\sqrt{N(|\beta_j|^2+|\beta_k|^2)}}.
\end{aligned}
\end{equation}
From the above derivations, we find that $N \mathbf{a}_j^\mathrm{H}$ is the MRT precoding vector of direction $\mathbf{a}_j$, and $(1-D_{k,j}^N)/(1-D_{k,j}) \mathbf{a}_k^\mathrm{H}$ is the MRT precoding vector of direction $\mathbf{a}_k$, which means that the ZF precoding vector of $\mathbf{a}_j$ is a combination of MRT precoding vectors.

Furthermore, the interference caused by beam $N \mathbf{a}_j^{\mathrm{H}}$ in direction $ \mathbf{a}_k$ is $\mathbf{a}_k N \mathbf{a}_j^{\mathrm{H}}=N (1-D_{k,j}^N)/(1-D_{k,j})$.
Moreover, the interference caused by beam $(1-D_{k,j}^N)/(1-D_{k,j}) \mathbf{a}_k^\mathrm{H}$ in direction $ \mathbf{a}_k$ is $\mathbf{a}_k (1-D_{k,j}^N)/(1-D_{k,j}) \mathbf{a}_k^\mathrm{H}=N (1-D_{k,j}^N)/(1-D_{k,j})$, which means the interference beam generated by the MRT beam of the served user is canceled by an MRT beam that is aligned with the interfered user at the same amplitude and antiphase as the interference.

In other words, when ZF sends a signal in the direction $\mathbf{a}_j$, it sends a cancellation signal in the direction $\mathbf{a}_k$ simultaneously so that the signal leaked in the direction $\mathbf{a}_k$ is canceled to zero by $\mathbf{a}^\mathrm{H}_k$.
However, the cancellation of the interference comes at a price, and the cancellation beam $(1-D_{k,j}^N)/(1-D_{k,j}) \mathbf{a}_k^\mathrm{H}$ will also cause attenuation in the served user signal:
\begin{equation}\label{Omega1}
\begin{aligned}
    \Omega(t_{j,k})=\mathbf{a}_j\frac{(1-D_{k,j}^N)}{(1-D_{k,j})} \mathbf{a}_k^\mathrm{H}
    &=\frac{1-\cos(N\pi t_{j,k})}{1-\cos(\pi t_{j,k})}.
\end{aligned}
\end{equation}
Now we can infer the following: When two directions $\mathbf{a}_j,\mathbf{a}_k$ are too close to each other, the MRT beam of $\mathbf{a}_j$ will greatly interfere with $\mathbf{a}_k$; in turn, the cancellation signal will also cause great signal attenuation for $\mathbf{a}_j$.
When the directions are far away, the attenuation is negligible.
\begin{figure}[!t]
\centering
\includegraphics[width=3.4in]{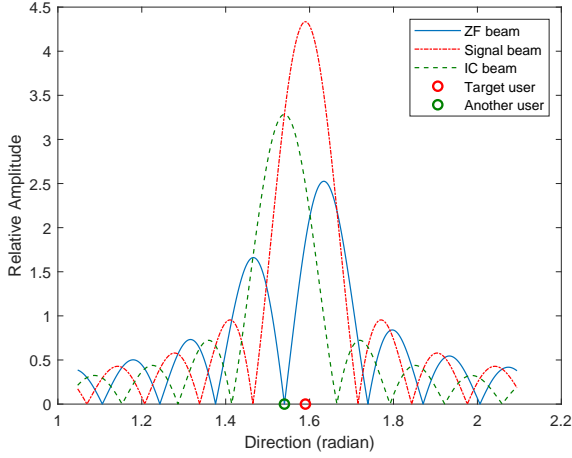}
\caption{Combination of ZF beams when the angular spacing of users is small.}
\label{zf_bclose}
\end{figure}

\begin{figure}[!t]
\centering
\includegraphics[width=3.4in]{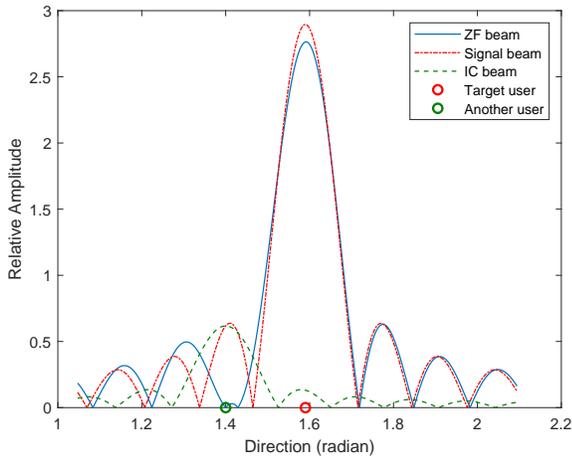}
\caption{Combination of ZF beams when the angular spacing of users is large.}
\label{zf_bfar}
\end{figure}

Then, we focus on the effect of the angular spacing of users on ZF beams.
In Fig. \ref{zf_bclose}, two users are close to each other, so we can see that the ZF beam combined with the signal beam and interference cancellation (IC) beam is much smaller than the signal beam.
The IC beam not only weakens the peak value of the ZF beam but also changes the energy distribution and increases the energy leakage.
In addition, the peak value of the ZF beam shifts from the target user position, which seriously affects the amplitude of the user's received signal.
These are the costs of interference cancellation.
However, in Fig. \ref{zf_bfar}, the ZF beam is almost equal to the signal beam, which reveals that ZF performs well when there is sufficient angular spacing between users.

\begin{figure}[!t]
\centering
\includegraphics[width=3.4in]{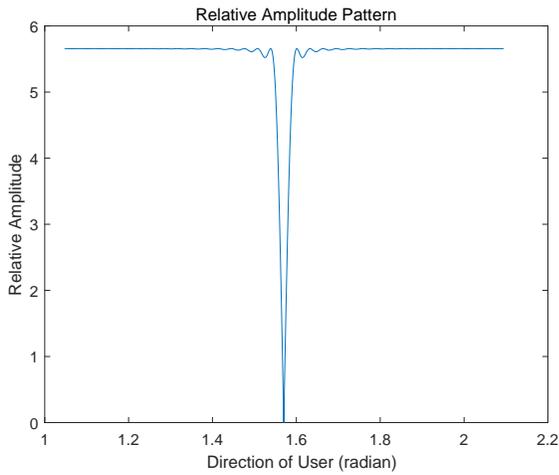}
\caption{Amplitude of the ZF beam of a user at a fixed angle while the angle of another user varies along the x-axis.}
\label{zf_am}
\end{figure}

In Fig. \ref{zf_am}, the angle of one user is fixed at $\frac{1}{2}\pi$, and the x-axis represents the varying angle of another user.
However, the amplitude now represents the signal amplitude of the target user.
Different from MRT precoding, the interference in ZF is nearly zero, but the amplitude of the target user decays sharply as the other user moves close to the target user: if the angular spacing between other users and the target user is too small, the performance of the system will decline rapidly.

The sufficient angular spacing could be set as $2/N$, which is the first zero point of $\Omega(t)$ according to \ref{Sec4C} and also half the width of the main lobe of Signal beam in Fig. \ref{zf_bclose} and \ref{zf_bfar}.

\subsubsection{ZF of multiuser systems}
Similar to the situation of a pair of users, a multiuser ZF beam can still be decomposed into a linear combination of MRT beams.
Set $\mathbf{J}=\mathbf{H}\mathbf{H^{\mathrm{H}}}$; then, the adjoint matrix of $\mathbf{J}$ is $\mathrm{adj}(\mathbf{J})$, and $J_{i,j}$ is the algebraic remainder of the element in row $i$ and column $j$ of $\mathbf{J}$, which is also the element in row $j$ and column $i$ of $\mathrm{adj}(\mathbf{J})$, we have
 
\begin{equation}\label{zf_comb_mrt}
    \mathbf{p}'_j=\frac{1}{\det(\mathbf{J})}(J_{j,1}\bar{\beta_1}\mathbf{a}_1^{\mathrm{H}}+J_{j,2}\bar{\beta_2}\mathbf{a}_2^{\mathrm{H}}+\dots+J_{j,K}\bar{\beta_K}\mathbf{a}_K^{\mathrm{H}}).
\end{equation}
The approximation and calculation of the determinant is a difficult problem in the field of mathematics, so we consider using the conclusion of \ref{pairZF} to approximate the ZF precoding vector.
To facilitate the discussion, without loss of generality, the users are already sorted by angle
\begin{equation}
    \theta_1<\theta_2<\ldots<\theta_K.
\end{equation}

From \ref{pairZF} we know that with the increase of angle space, the cost of interference elimination decreases rapidly.
At the same time, to avoid paying too much to interference elimination, the minimum angle space could not be too small for any user pair.
This means, the angular spacing between adjacent users cannot be too small.
This also makes the angular spacing between other users except adjacent users maintain a large level, so that the cost of interference elimination can be ignored.
Therefore, ZF vector of user $j, 2 \leq j \leq K-1$ could be approximate as
\begin{equation}\label{zf_aprox0}
    \mathbf{p}'_j \approx \frac{1}{\zeta_j}(J_{j,j-1}\bar{\beta}_{j-1}\mathbf{a}_{j-1}^{\mathrm{H}}+J_{j,j}\bar{\beta}_j\mathbf{a}_j^{\mathrm{H}}+J_{j,j+1}\bar{\beta}_{j+1}\mathbf{a}_{j+1}^{\mathrm{H}}),
\end{equation}
where $\zeta_j$ is for power control.
Since the orthogonality between vectors is independent of the module, we leave the module value problem for later.
In fact, since the angular spacing between user $j-1$ and $j+1$ is large enough, the interaction between the two users is small enough to be ignored. 
Therefore, the conclusion of pair user ZF of \eqref{pair_user_zf} can be applied, then $\mathbf{p}'_j$ could be further approximated as
\begin{equation}\label{zf_aprox1}
    \mathbf{p}'_j \approx \frac{1}{\zeta_j}(N \mathbf{a}_j^\mathrm{H} - \frac{1-D_{j-1,j}^N}{1-D_{j-1,j}}\mathbf{a}_{j-1}^{\mathrm{H}}-\frac{1-D_{j,j+1}^N}{1-D_{j,j+1}}\mathbf{a}_{j+1}^{\mathrm{H}}).
\end{equation}
After normalizing its power, it can be written as 
\begin{subequations}\label{zf_aprox}
    \begin{align}
        \zeta_j & = \left \| N \mathbf{a}_j^\mathrm{H} - \frac{1-D_{j-1,j}^N}{1-D_{j-1,j}}\mathbf{a}_{j-1}^{\mathrm{H}}-\frac{1-D_{j,j+1}^N}{1-D_{j,j+1}}\mathbf{a}_{j+1}^{\mathrm{H}}\right \|_F\\
        \widetilde{\mathbf{p}}_j & = \frac{1}{\zeta_j}\left ( N \mathbf{a}_j^\mathrm{H} - \frac{1-D_{j-1,j}^N}{1-D_{j-1,j}}\mathbf{a}_{j-1}^{\mathrm{H}}-\frac{1-D_{j,j+1}^N}{1-D_{j,j+1}}\mathbf{a}_{j+1}^{\mathrm{H}}\right )
    \end{align}
\end{subequations}

\subsubsection{MMSE}
The MMSE precoding method, which is a compromise of MRT and ZF, is similar to both MRT and ZF: when noise $\sigma^2$ is small, MMSE will act like ZF; when noise $\sigma^2$ is too large to be ignored, MMSE will act like MRT.
In most cases, the above ZF rules are also applied to MMSE.
However, MMSE precoding has its characteristics.
When MMSE eliminates interference between users, it considers the impact of noise and balances the values of noise and interference.
Compared with ZF, it increases a user's signal strength to combat noise instead of blindly seeking zero interference.

\subsection{Free space multiuser signal strength}
\begin{figure*}[!t]
\centering
\subfigure[MRT]{\includegraphics[width=2.35in]{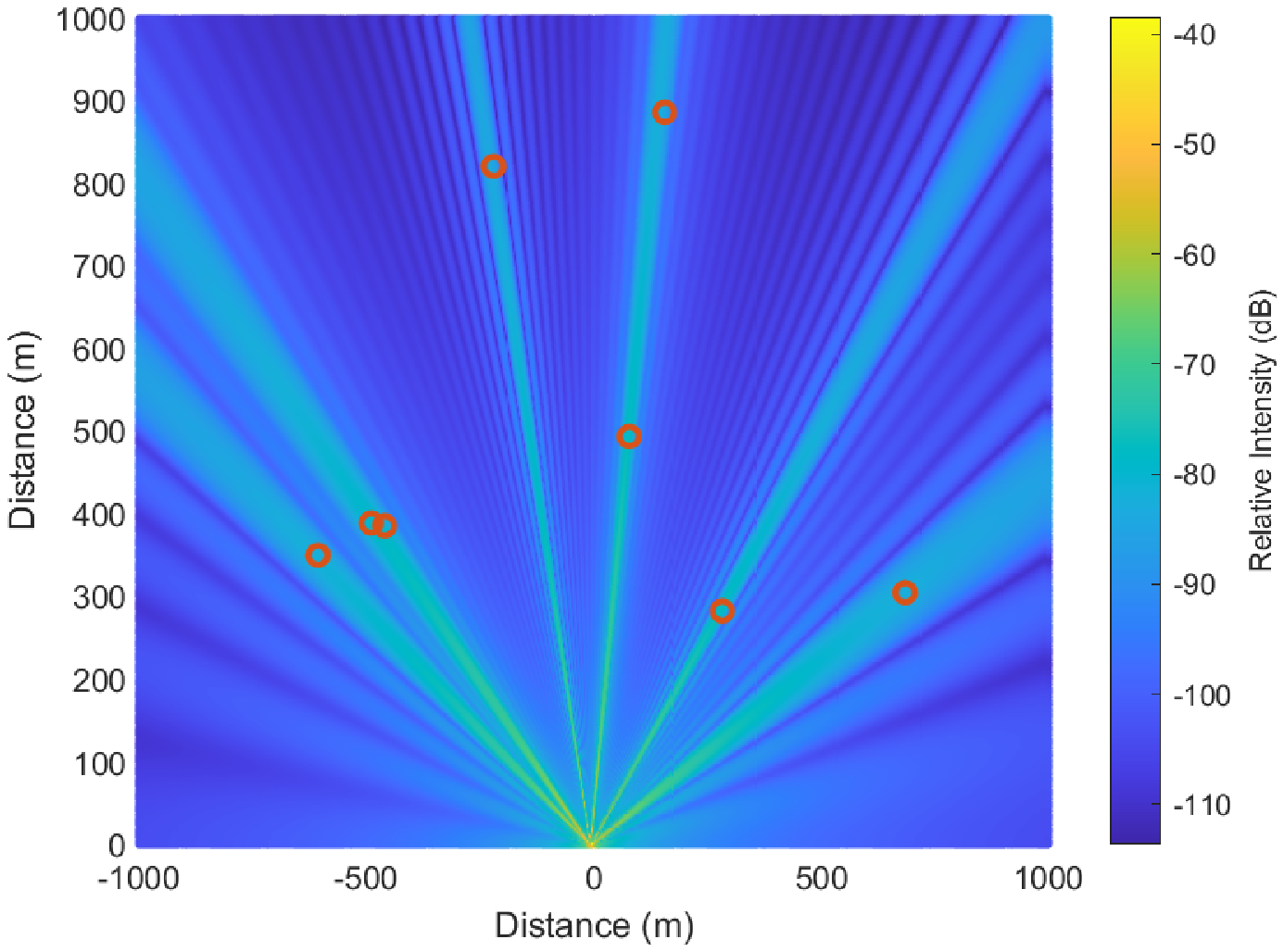}%
\label{BMRT}}
\subfigure[ZF]{\includegraphics[width=2.35in]{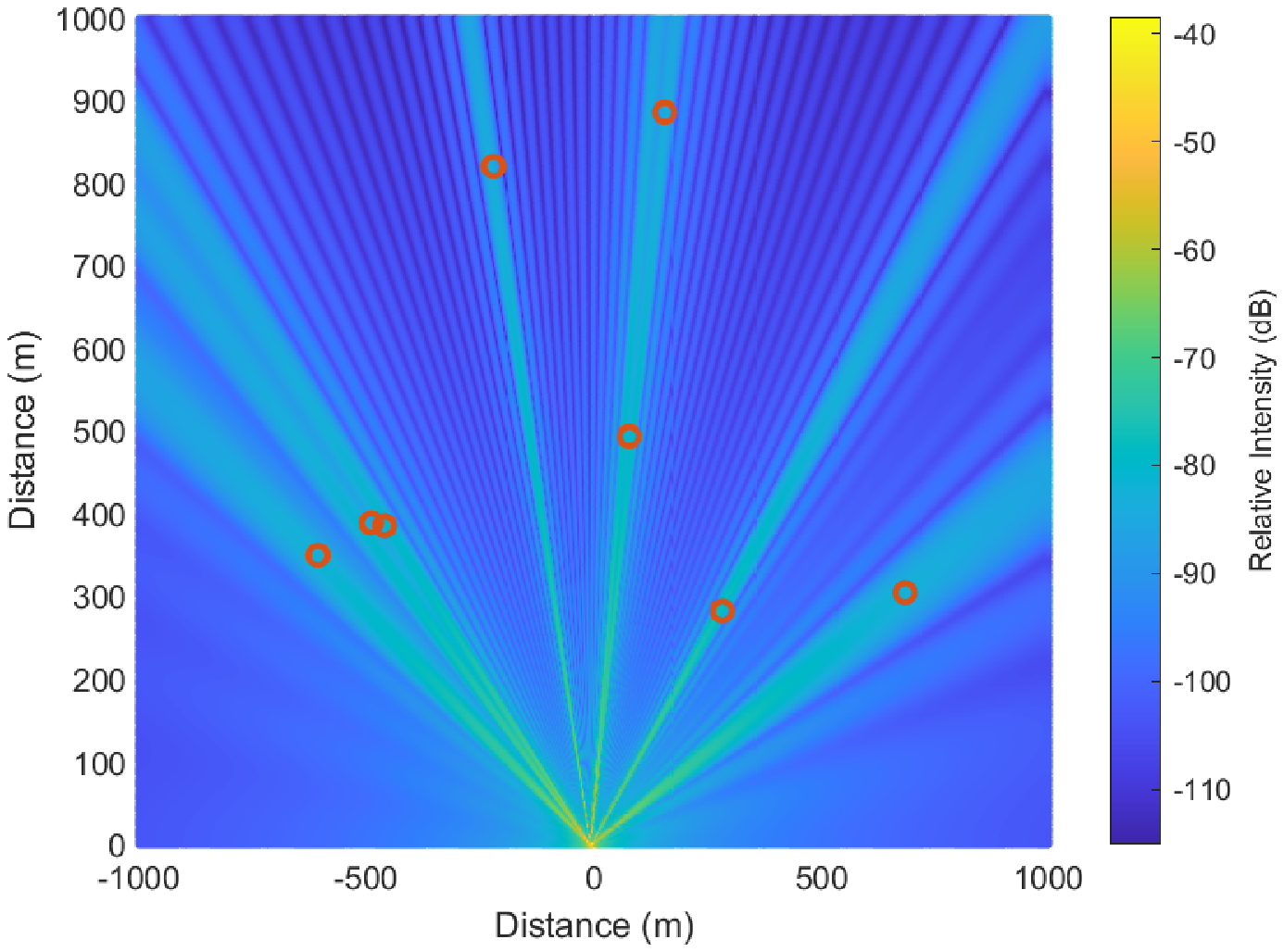}%
\label{BZF}}
\subfigure[MMSE]{\includegraphics[width=2.35in]{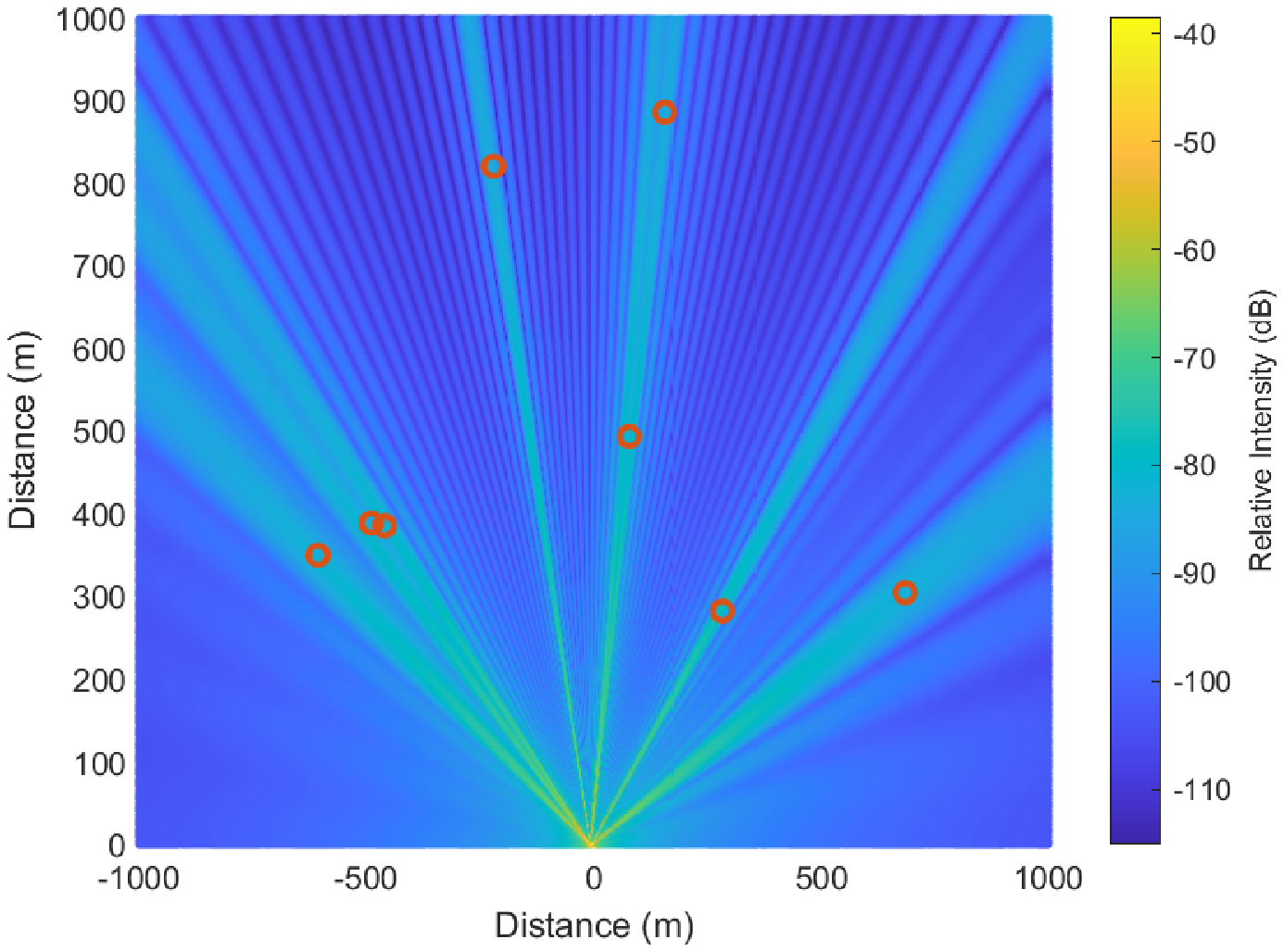}%
\label{BMMSE}}
\caption{Distribution of radiation intensity for multiuser linear precoding.}
\label{B-LP}
\end{figure*}

To show the beamforming effect more clearly, we set up an ideal transmission environment, which means that only the path loss and steering vector are considered.
Fig. \ref{B-LP} simulates a rectangular cell: The base station is located at the origin, the ULA plane is parallel to the x-axis, the black circle in the figure represents the user's position, and the base station uses linear precoding to serve $K=8$ users in the cell at the same time.
The colors in the graph represent the intensity of all the received signals at each location and do not distinguish between signals and interference.

Fig. \ref{B-LP} shows that multiple narrow areas with strong signals, which are beams, originating from the base station accurately point to multiple users in the cell.
For a user located far away from other users, one beam can be sent by the base station.
However, for adjacent users, the MRT cannot distinguish them well. 
For example, the beams of three adjacent users located in the upper left corner overlap each other, which provides the potential for interference.
For ZF and MMSE, it can be seen that even if the angular spacing of users is small, different beams can be distinguished.
Now, the principle of precoding in mmWave massive MIMO is revealed:
Linear precoding concentrates signal energy and reduces interference by aligning the beams.

\subsection{Angular Multiplexing Gain and Group Gain}
\subsubsection{Angular Multiplexing Gain}
In the previous part of this section, we discussed the beamforming effect of precoding and the impact of paired user angular spacing on precoding performance.
It is not difficult to find that this effect will also have a great impact on the sum-rate of the system.
Therefore, in the following, the improvement in the sum-rate of a system brought by precoding is called the angular multiplexing gain.
The improvement in the sum-rate of a system brought by a reasonable angle distribution of users is called the angular multiplexing gain.

\subsubsection{Grouping Gain}
The above discussion shows that when the number of users is large, the performance loss of ZF and MMSE precoding is very serious.
Therefore, to cope with a large number of users, that is, when the number of users in the cell is larger than the number of antennas, a mix of two or more multiple access systems to serve users must be considered.
The sum-rate of the system that applies ZF precoding is almost reduced to zero when the number of users approaches the number of antennas.
Thus, when we group users, the distance (or angular distance) between users will be naturally widened, which will alleviate the detriment caused by large numbers of users.
Therefore, compared with the case of no grouping, user grouping will bring very large gains to the system when the number of users is large, especially when the users are located close together.

\section{Problem Formulation}\label{Sec4}
For multiple access systems implemented by precoding, especially MRT and MMSE, precoding itself can achieve any number of data streams, which is the key to multiple access systems.
However, as the number of users grows, the sum-rate of the system will reach a bottleneck.
At this time, the time division or frequency division is combined and grouped according to certain rules based on the space or angle to effectively improve the sum-rate of the system.

In this section, we formulate the optimization problem and use the result of Section \ref{Sec3} to turn it into an approximation problem that is easier to solve.

\subsection{Problem of Maximizing the sum-rate }
The problem of maximizing the sum-rate can be stated as
\begin{subequations}\label{MAXrate}
\begin{align}
  \mathcal{P}_0: \mathop {\max }\limits_{\mathcal{U}_1,\mathcal{U}_2,...,\mathcal{U}_G} \quad
  & \frac{1}{G}\sum_{g=1}^{G}\sum_{k\in\mathcal{U}_g} R_k \\
  {\rm{s.t.}} \quad &\mathcal{U}=\{1,2,...,K\} \label{UK} \\ 
    &\mathcal{U}=\mathcal{U}_1\cup\mathcal{U}_2\cup...\cup\mathcal{U}_G \label{UU} \\  
    &\forall i\neq k,\mathcal{U}_i\cap\mathcal{U}_k=\emptyset. \label{UA}
\end{align}
\end{subequations}
This grouping problem is a typical NP-hard problem, which is almost impossible to directly solve.
Some approximate laws are needed to solve this problem.

\subsection{The impact of angular spacing on precoding performance}\label{R(t)}
This subsection takes advantage of the beamforming effect of linear precoding to simplify the optimization problem.

According to Section \ref{Sec3}, we know that in an mmWave Massive MIMO system, $\mathrm{SINR}_k$ is highly correlated with the angular spacing.
For example, in the MRT precoding method, the signal power is relatively constant, and the interference between user $k$ and $j$ is determined by $|\psi _{k}-\psi _{j}|=\frac{d}{\lambda}|\cos \theta_k-\cos \theta_j|=\frac{d}{\lambda}|t_{k,j}|$.
Especially when $\frac{1}{6}\pi \le \theta_k, \theta_j \le \frac{5}{6}\pi$, we have $|t_{k,j}| \propto |\theta_k-\theta_j|$.
Moreover, interuser interference mainly occurs between adjacent users.
Therefore, the angular space between adjacent users has a decisive influence on the performance of the system.

Due to the better performance and the smoother relationship between the signal and angle, we choose the ZF precoding method for research.
When grouping and ZF precoding are adopted, the rate of user $k$ could be rewritten as 
\begin{equation}\label{kGRate}
    R_k=\log_2\left( 1+\frac{p_k\|\mathbf{h}_k\mathbf{p}^{\mathcal{U}_i}_k\|_F^2}
    {\sigma^2}\right),
\end{equation}
where $k \in \mathcal{U}_i$, $p_k$ is the beam power to user $k$, and $\mathbf{p}^{\mathcal{U}_i}_k$ is the ZF precoding vector of user $k$ in group $\mathcal{U}_i$, which means $\mathbf{p}^{\mathcal{U}_i}_k$ is calculated by using the channel matrix composed of users in $\mathcal{U}_i$.
For the sake of brevity, we use $\mathbf{p}_k$ to represent $\mathbf{p}^{\mathcal{U}_i}_k$.

Then the received signal of user $j$ under ZF precoding is shown as \eqref{reci_sig} at the top of the Page \pageref{reci_sig}.
\begin{figure*}
    \begin{equation}\label{reci_sig}
        p_j\|\mathbf{h}_j\mathbf{p}^{\mathcal{U}_i}_j\|_F^2 \approx p_j\|\mathbf{h}_j\widetilde{\mathbf{p}}_j\|_F^2 = \frac{p_j}{\zeta^2_j}\left ( N \mathbf{a}_j \mathbf{a}_j^\mathrm{H} - \frac{1-D_{j-1,j}^N}{1-D_{j-1,j}} \mathbf{a}_j \mathbf{a}_{j-1}^{\mathrm{H}}-\frac{1-D_{j,j+1}^N}{1-D_{j,j+1}} \mathbf{a}_j \mathbf{a}_{j+1}^{\mathrm{H}}\right )^2.
    \end{equation}
    \noindent\rule[0.25\baselineskip]{\textwidth}{0.8pt}
\end{figure*}
Using $\Omega(t)$ in \eqref{Omega1}, \eqref{reci_sig} can be further reduced to
\begin{equation}
    p_k\|\mathbf{h}_k\widetilde{\mathbf{p}}_k\|_F^2 = \frac{p_k}{\zeta^2_k}|\beta_k|^2\left ( N^2 - \Omega(t_{k_{-},k})-\Omega(t_{k,k_{+}}) \right )^2,
\end{equation}
where $k_{+}$ denotes the larger element that is closest to $K$ in the same group, while $k_{-}$ denotes the smaller element that is closest to K in the same group.
After that, \eqref{kGRate} could be rewritten as 
\begin{equation}\label{Rate_aprox}
    R_k \approx \log_2\left( 1+\Gamma_{k}\left ( N^2 - \Omega(t_{k_{-},k})-\Omega(t_{k,k_{+}}) \right )^2\right),
\end{equation}
where $\Gamma_{k}=(p_k|\beta_k|^2)/(\zeta^2_k\sigma^2)$.
Generally speaking, the array gain of massive MIMO is high, and most users are in the high SNR region,
\begin{subequations}
    \begin{align}
        R_k & \approx \log_2\left(\Gamma_{k}\left ( N^2 - \Omega(t_{k_{-},k})-\Omega(t_{k,k_{+}}) \right )^2\right)\\
        & = \log_2(\Gamma_{k}N^4)+2\log_2\left ( 1 - \frac{1}{N^2}(\Omega(t_{k_{-},k})+\Omega(t_{k,k_{+}})) \right )\\
        & \approx  \log_2(\Gamma_{k}N^4)-\frac{2}{N^2}(\Omega(t_{k_{-},k})+\Omega(t_{k,k_{+}}))\\
        & = \widetilde{R}_k.
    \end{align}
\end{subequations}
Therefore, the optimization objective can be approximated as \eqref{P1_obj} at the top of the Page \pageref{P1_obj}.
\begin{figure*}
    \begin{subequations}
        \begin{align}
            & \frac{1}{G}\sum_{g=1}^{G}\sum_{k\in\mathcal{U}_g} R_k\approx \frac{1}{G}\sum_{g=1}^{G}\sum_{k\in\mathcal{U}_g} \widetilde{R}_k = \frac{1}{G}\sum_{g=1}^{G}\sum_{k\in\mathcal{U}_g}\left ( \log_2(\Gamma_{k}N^4)-\frac{2}{N^2}(\Omega(t_{k_{-},k})+\Omega(t_{k,k_{+}})) \right )\\
            & = \frac{1}{G}\sum_{g=1}^{G}\sum_{k\in\mathcal{U}_g} \log_2(\Gamma_{k}N^4)-\frac{2}{GN^2}\sum_{g=1}^{G}\sum_{k\in\mathcal{U}_g} (\Omega(t_{k_{-},k})+\Omega(t_{k,k_{+}})).\label{P1_obj}
        \end{align}
    \end{subequations}
    \noindent\rule[0.25\baselineskip]{\textwidth}{0.8pt}
\end{figure*} 

To facilitate the discussion, without loss of generality, the users are already sorted by angle
\begin{equation}
    \theta_1<\theta_2<\ldots<\theta_K.
\end{equation}
Set $u^i_m \in \{1,2,...,K\}$ as the $m$-th user in group $i$, $M$ is the number of elements in $\mathcal{U}_i$,
\begin{equation}
    \mathcal{U}_i = \{ u^i_1,u^i_2,\ldots,u^i_M \}.
\end{equation}
We set the elements in the group to be arranged in an increasing order, which means that $u^i_1<u^i_2<\ldots<u^i_M$, and the corresponding angles of $\mathcal{U}_i$ are also arranged
\begin{equation}
    \theta_{u^i_1}<\theta_{u^i_2}<\ldots<\theta_{u^i_M}.
\end{equation}

In addition, since the boundary elements, which are the elements with the smallest and largest angle in the group, lack one of the neighboring elements, we have
\begin{equation}
    \sum_{g=1}^{G}\sum_{k\in\mathcal{U}_g} -(\Omega(t_{k_{-},k})+\Omega(t_{k,k_{+}}))=2\sum_{g=1}^{G}\sum_{k=1}^{M-1} -\Omega(t_{u^g_k,u^g_{k+1}})
\end{equation}

For equal power allocation, 
\begin{equation}
    \frac{1}{G}\sum_{g=1}^{G}\sum_{k\in\mathcal{U}_g} \log_2(\Gamma_{k}N^4)
\end{equation}
will hardly change.
Therefore, the problem can be approximated as
\begin{subequations}\label{MAXtheta}
    \begin{align}
      \mathcal{P}_1: \mathop {\max }\limits_{\mathcal{U}_1,\mathcal{U}_2,...,\mathcal{U}_G} \quad
      & \frac{1}{G}\sum_{g=1}^{G}\sum_{k=1}^{M-1} -\Omega(t_{u^g_k,u^g_{k+1}}) \\
      {\rm{s.t.}} \quad &\eqref{UK},\eqref{UU},\eqref{UA},
    \end{align}
\end{subequations}

\subsection{Maximizing the Sum of a Certain Revenue Function for the Angular Spacing}\label{Sec4C}
The properties of $\Omega(t_{u^g_k,u^g_{k+1}})$ are the key to the optimization problem.
Next, we will get rid of the subscripts of $t_{u^g_k,u^g_{k+1}}$ for a moment and study the properties of this function alone:
\begin{subequations}
    \begin{align}
        \Omega(t) = & \frac{1-\cos(N\pi t)}{1-\cos(\pi t)}\\
        \frac{\mathrm{d}\Omega(t)}{\mathrm{d}t} = & \frac{\pi N \sin{ (N \pi t)} (1-\cos{(\pi t)})}{(1-\cos{(\pi t)})^2} \nonumber\\
        &-\frac{\sin{(\pi t)}(1-\cos{(1-\cos{(N \pi t)})})}{(1-\cos{(\pi t)})^2}.
    \end{align}    
\end{subequations}
When $0<t<\frac{2}{N}$, we have $\Omega(t)>0,\mathrm{d}\Omega(t)/\mathrm{d}t<0$.
The analysis of the second derivative $\mathrm{d}^2\Omega(t)/\mathrm{d}t^2$ could be done by simulation.
\begin{figure*}[!t]
    \centering
    \subfigure[$\Omega(t)$]{\includegraphics[width=2.35in]{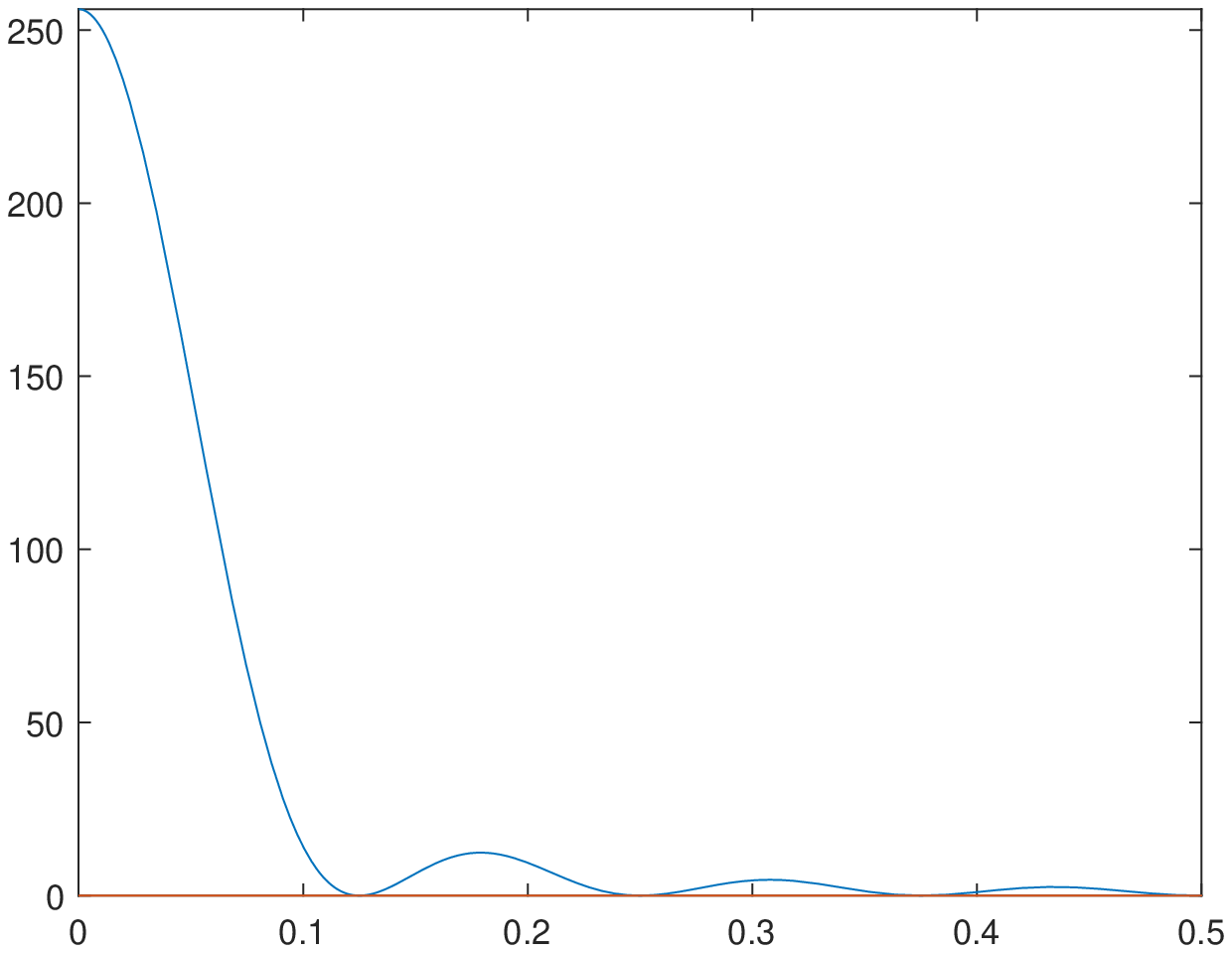}%
    \label{Ot}}
    \subfigure[$\frac{\mathrm{d}\Omega(t)}{\mathrm{d}t}$]{\includegraphics[width=2.35in]{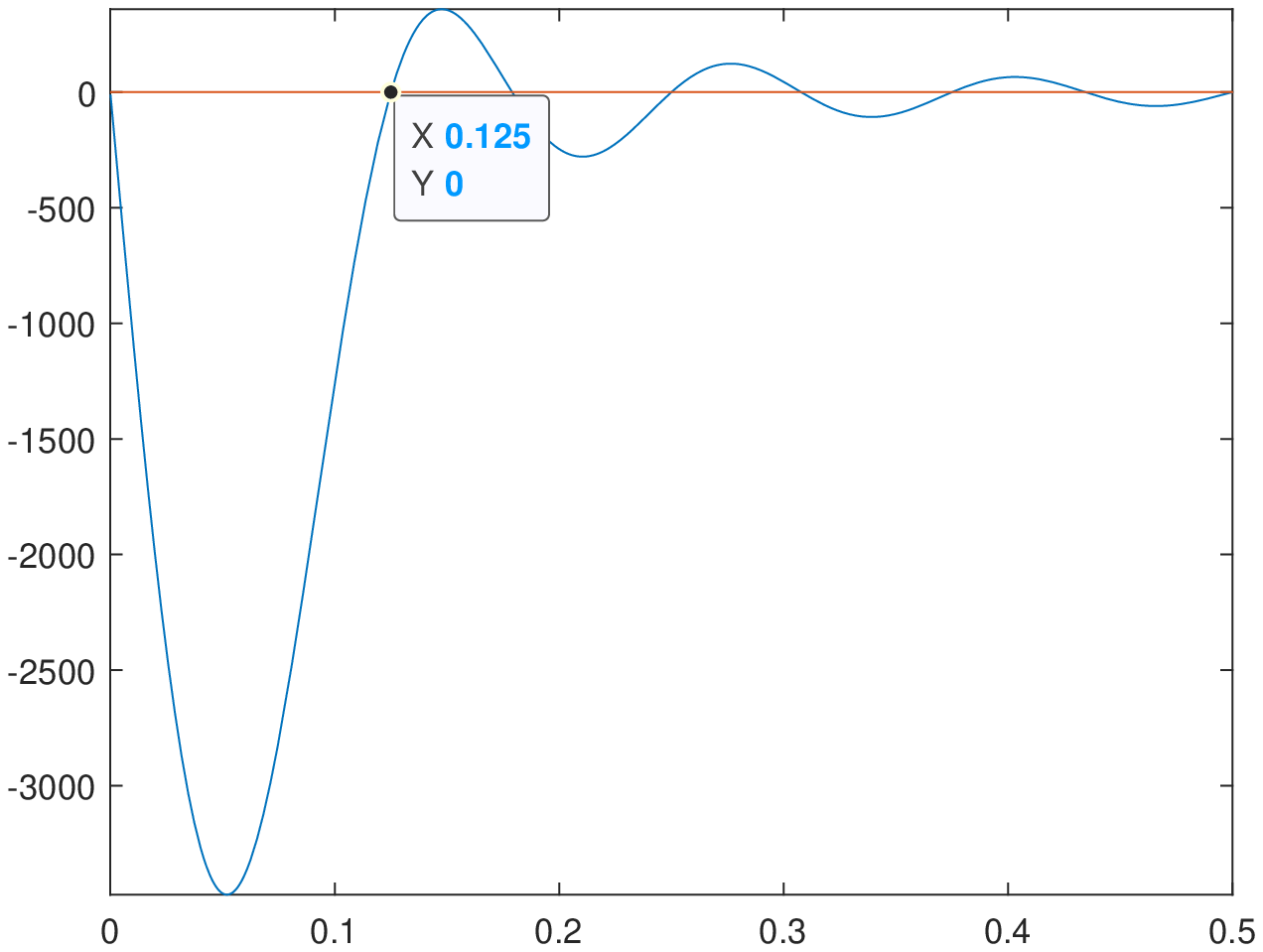}%
    \label{dOt}}
    \subfigure[$\frac{\mathrm{d}^2\Omega(t)}{(\mathrm{d}t)^2}$]{\includegraphics[width=2.35in]{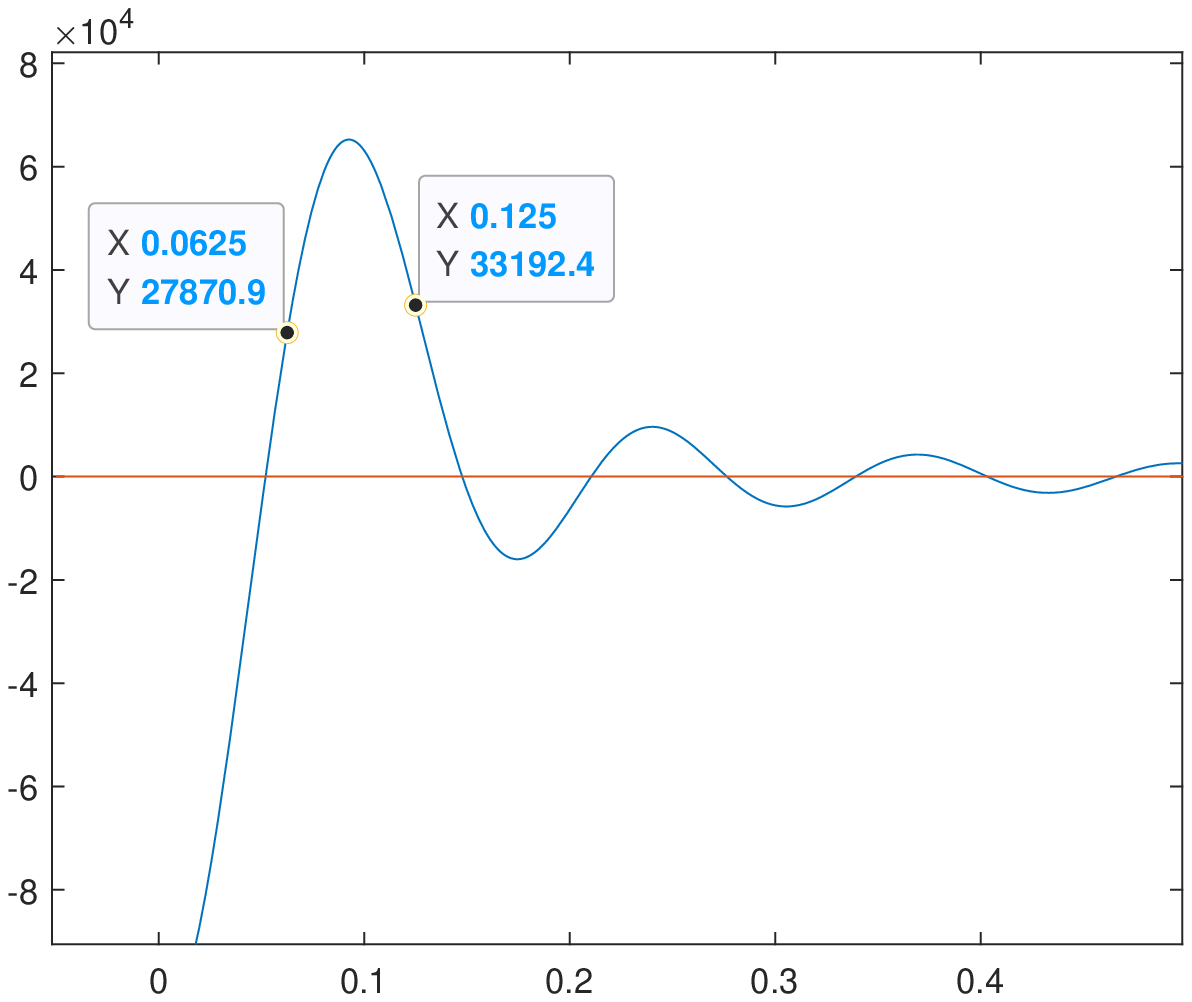}%
    \label{ddOt}}
    \caption{Function $\Omega(t)$ and its first-order and second-order derivatives when $N=16$.}
    \label{fig_sim}
\end{figure*}
As shown in  Fig. \ref{fig_sim} at the top of Page. \pageref{fig_sim},  we have $\mathrm{d}^2\Omega(t)/(\mathrm{d}t)^2>0, 1/N<t<2/N$.
According to the discussion in Section \ref{Sec3}, the larger the angular spacing $t$, the better performance of precoding we will get. 
As also shown in Fig. \ref{Ot}, when $t$ is close to $2/N$, the performance of precoding achieves the highest performance.
At the same time, if $t$ is too large, the group could not contain too much users, which will in turn cause performance reduction.
Therefore, the $t$ better be a little less than $2/N$.

In this case, we define a function class $g(t)$, where $g'(t)>0,g''(t)<0,g(t)=g(-t)$.
In fact, $-\Omega(t)$ when $1/N<t<2/N$ satisfies the definition of $g(t)$, what is true for $g(t)$ should also be true for $-\Omega(t)$.

Finally, the optimization problem can be approximated as
\begin{subequations}\label{Finally}
\begin{align}
  \mathcal{P}_2: \mathop {\max }\limits_{\mathcal{U}_1,\mathcal{U}_2,...,\mathcal{U}_G} \quad
  & \mathrm{Obj}=\frac{1}{G}\sum_{g=1}^{G}\sum^{M-1}_{k=1} g(\theta_{u^g_{k+1}}-\theta_{u^g_k}) \\
  {\rm{s.t.}} \quad &\eqref{UK},\eqref{UU},\eqref{UA},
\end{align}
\end{subequations}
where $\theta_{u^g_{k+1}}-\theta_{u^g_k} \approx t_{u^g_{k+1},u^g_{k}},\forall x, g(x)=g(-x)$ and $\forall x>0,g'(x)>0, g''(x)<0$.

\section{Proposed Angular Spacing Equalization Grouping Algorithm}\label{Sec5}
In this section, we derive the optimal solution to Problem $\mathcal{P}_2$ and the corresponding ASEG algorithm.
Then, the comparison algorithms are introduced.

\subsection{Angular Spacing Equalization Grouping}
It can be seen from the above research that grouping methods are beneficial when the number of users is large, so we first make the large-scale user grouping assumption that each group has multiple users (at least greater than or equal to 3).
This assumption is not hard to satisfy.
We generally adopt the strategy of equal-member grouping.

In this case, it is difficult to design the optimal grouping algorithm directly, but we can analyze the best grouping results in some cases:

{\bf{Proposition 1:}} Within a group (excluding the boundaries), there cannot be more than one element whose value is between a group of adjacent elements in any other group.
Mathematically, for group $\mathcal{U}_i = \{ u^i_1,u^i_2,\ldots,u^i_M \}$ and group $\mathcal{U}_j = \{ u^j_1,u^j_2,\ldots,u^j_M \}$, if there are $u^j_k$ and $u^i_l $ such that $u^i_l<u^j_k<u^j_{k+1}<u^i_{l+1}$, then there exists a better grouping than this grouping.

{\bf{Proof}:}
Exchange the elements before $u^j_k$ and $u^j_k$ in group $\mathcal{U}_j$ and the elements before $u^i_l$ and $u^i_l$ in group $\mathcal{U}_i$; namely,
\begin{subequations}
\begin{align}
\mathcal{U}'_i &= \{ u^j_1,u^j_2,\ldots,u^j_k,u^i_{l+1}\ldots,u^i_M \} \\
\mathcal{U}'_j &= \{ u^i_1,u^i_2,\ldots,u^i_l,u^j_{k+1}\ldots,u^j_M \}.
\end{align}
\end{subequations}
Suppose the objective function before the exchange is $\mathrm{Obj}$, and the objective function after the exchange is $\mathrm{Obj'}$; then, we have
\begin{equation}\label{Rule1-1}
\begin{aligned}
    &\mathrm{Obj}-\mathrm{Obj'}=\\
    &g(\theta_{u^j_{k+1}}-\theta_{u^j_{k}})+g(\theta_{u^i_{l+1}}-\theta_{u^i_{l}})\\
    &-[g(\theta_{u^j_{k+1}}-\theta_{u^i_{l}})+g(\theta_{u^i_{l+1}}-\theta_{u^j_{k}})].
\end{aligned}
\end{equation}
\eqref{Rule1-1} can be rewritten as \eqref{P1A} at the top of the Page \pageref{P1A}.
Notice that $g''(x)<0$, so we have \eqref{P1B}, which means \eqref{P1C}.
Then, we have $\mathrm{Obj}<\mathrm{Obj'}$; that is, $\mathcal{U}'_i$ and $\mathcal{U}'_j$ are better groupings
$\hfill\blacksquare$
\begin{figure*}
    \begin{subequations}
        \begin{align}
            & \mathrm{Obj}-\mathrm{Obj'}=\int^{\theta_{u^j_{k}}-\theta_{u^i_{l}}}_{0}[g'(x+\theta_{u^j_{k+1}}-\theta_{u^j_{k}}+\theta_{u^i_{l+1}}-\theta_{u^j_{k+1}})-g'(x+\theta_{u^j_{k+1}}-\theta_{u^j_{k}})]dx \label{P1A}\\
            & [g'(x+\theta_{u^j_{k+1}}-\theta_{u^j_{k}}+\theta_{u^i_{l+1}}-\theta_{u^j_{k+1}})-g'(x+\theta_{u^j_{k+1}}-\theta_{u^j_{k}})]<0\label{P1B}\\
            & \int^{\theta_{u^j_{k}}-\theta_{u^i_{l}}}_{0}[g'(x+\theta_{u^j_{k+1}}-\theta_{u^j_{k}}+\theta_{u^i_{l+1}}-\theta_{u^j_{k+1}})-g'(x+\theta_{u^j_{k+1}}-\theta_{u^j_{k}})]dx<0\label{P1C}
        \end{align}
    \end{subequations}
    \noindent\rule[0.25\baselineskip]{\textwidth}{0.8pt}
\end{figure*}

{\bf{Proposition 2:}} The other groups cannot have multiple elements larger than the boundary of one side of a certain group.
Mathematically, if there is $ u^{i}_{k}$ such that $u^j_M<u^i_{k}<u^i_{k+1}$, then there exists a better grouping than this grouping.

{\bf{Proof}:} Transfer the element $u^i_{k+1}$ in group $\mathcal{U}_i$ to group $\mathcal{U}_j$, and the objective function before and after the exchange can be expressed as
\begin{equation}
\begin{aligned}
    &\mathrm{Obj}-\mathrm{Obj'}\\
    &=g(\theta_{u^i_{k+1}}-\theta_{u^i_{k}})-g(\theta_{u^i_{k+1}}-\theta_{u^i_M}).
\end{aligned}
\end{equation}
Obviously, $(\theta_{u^i_{k+1}}-\theta_{u^i_M})>(\theta_{u^i_{k+1}}-\theta_{u^i_{k}})$, and $g'(x)>0$.
Then, we have $\mathrm{Obj}<\mathrm{Obj'}$, i.e., we have a better grouping.
$\hfill\blacksquare$

{\bf{Proposition 3:}} Suppose a complete set consists of any two groups; then, the adjacent elements in it must belong to different groups.
Mathematically, for $\mathcal{U'}=\{\mathcal{U}_i,\mathcal{U}_j\}$, if $u^i_k$ is not the largest element of $\mathcal{U'}$, then $u^i_{k}+1 \in \mathcal{U}_j$; if $u^i_k$ is not the smallest element of $\mathcal{U'}$, then $u^i_{k}-1 \in \mathcal{U}_j$.

{\bf{Proof}:}
Now consider any two groups, for example,  $\mathcal{U}_i,\mathcal{U}_j$. Due to Proposition 1 and Proposition 2, we know that the elements of the two groups must be staggered elements in a complete set in increasing order.
That is, for any non-boundary element in $\mathcal{U}_i$, its adjacent elements in the complete set composed of $\mathcal{U}_i,\mathcal{U}_j$ must be elements of $\mathcal{U} _j$.
Otherwise, it will violate Proposition 1.
The elements in $\mathcal{U}_j$ also follow this rule.
The boundary elements also comply with the above rules.
The inner adjacent element of boundary elements in the complete set must be one of the elements in another set, and there are no elements on the outside of the boundary.
Thus, the numbers of elements in any two groups differ by no more than 1, and they are staggered in a complete set.
$\hfill\blacksquare$

{\bf{Proposition 4:}} For any adjacent elements in any group, there must be $G-1$ elements between them in a universal set.
Mathematically, $\forall u^i_k,1\leq k \leq M-1$, there exists only $G-1$ elements $\in \{u^i_k+1,u^i_k+2,...,u^i_k+G-1\}$ between $[u^i_k,u^i_{k+1}]$.

{\bf{Proof}:}
Due to Proposition 2 and Proposition 3, we might as well set $u^1_1=1$, that is, the first element of the first set of $\mathcal{U}_1$.
Guaranteed by Proposition 3, at this time, there must be a unique element between $u_{1}^{1}$ and $u_{2}^{1}$ in each other group, and they must be the $G-1$ elements between $[2, G]$.
If there are $t_1<t_2<\dots<t_T \in [u^1_1,u^1_2], T \geq 1$
and they do not belong to $[2, G]$, then in the remaining $G-1$ groups, there must be $G+T-1$ elements between $[u^1_1,u^1_2]$ and there must be at least one group that contains multiple elements between $[u^1_1,u^1_2]$, which violates Proposition 1.

Now, we consider any non boundary element $u^i_k$ of any group; 
there exist and only exist $G-1$ elements between $[u^i_{k-1},u^i_{k}]$, and the same between $[u^i_{k},u^i_{k+1}]$.
The proof of this conclusion is the same as the proof above, so it will not be repeated here.
$\hfill\blacksquare$

From Proposition 4, we know that the difference in the global sequence number for a certain group of adjacent elements is the number of groups $G$.
According to Proposition 2, the boundary elements of $G$ groups must be $\{1, 2, \dots, G\}$.
These $G$ elements, without loss of generality, are placed into the ${ \mathcal{U}_1, \mathcal{U}_2,\dots, \mathcal{U}_G}$ group; namely,
\begin{equation}
    1 \in\mathcal{U}_1,2 \in \mathcal{U}_2,\dots,G \in \mathcal{U}_G
\end{equation}

Then, according to the above grouping rules, we can determine a unique grouping method that satisfies all the above rules:
\begin{subequations}\label{ASEG-gro}
\begin{align}
    \mathcal{U}_1&=\{1,G+1,\dots,(M-1)G+1\}\\
    \mathcal{U}_2&=\{2,G+2,\dots,(M-1)G+2\}\\
    &\vdots\\
    \mathcal{U}_G&=\{G,2G,\dots,MG\}.
\end{align}
\end{subequations}
All the rules jointly determine a unique grouping fraction. 
Then, this grouping method is the optimal grouping method.

{\bf{Proposition 5:}} The optimal solution of Problem $\mathcal{P}_2$ is given by \eqref{ASEG-gro}.

Then, the ASEG algorithm can be constructed as Algorithm \ref{ASEG-alg}.

\begin{algorithm}
\caption{Angular Spacing Equalization Grouping (ASEG)}
\label{ASEG-alg}
\begin{algorithmic}[1]
\State{\textbf{Input:}} angle of all users $\mathcal{\theta}=[\theta_1,\theta_2,...,\theta_K]$
\State{apply a sorting algorithm, such as quicksort or mergesort, to $\mathcal{\theta}$}
\State{group users according to \eqref{ASEG-gro} into $\mathcal{U}_1,\mathcal{U}_2,...,\mathcal{U}_G$}
\State{\textbf{Output:} $\mathcal{U}_1,\mathcal{U}_2,...,\mathcal{U}_G$ }
\end{algorithmic}
\end{algorithm}

\subsection{Self-Evolution Genetic Algorithm for Comparison}
Genetic algorithms are widely used in grouping and selection problems \cite{MakkiS2018TWC, Wang2020TCOMM,Zhu2022LCOMM}.
Given the particularity of the grouping problem, we propose a self-evolution genetic algorithm (SEGA) based on the idea of a genetic algorithm.
SEGA is an improved genetic algorithm that fits the particularity of the grouping problem.

First, we need to return to Problem $\mathcal{P}_0$.
\begin{subequations}
\begin{align}
  \mathcal{P}_0: \mathop {\max }\limits_{\mathcal{U}_1,\mathcal{U}_2,...,\mathcal{U}_G} \quad
  & \frac{1}{G}\sum_{g=1}^{G}\sum_{k\in\mathcal{U}_g} R_k \\
  {\rm{s.t.}} \quad &\eqref{UK},\eqref{UU},\eqref{UA}.
\end{align}
\end{subequations}
where $\mathcal{U}_g$ is naturally coded.
Now, we redefine $\mathcal{U}_{g,e}$ as the $g$-th group of elite $e$ elements, which is a special group with the best gene.
Due to the uniqueness of users described by\eqref{UK},\eqref{UU} and \eqref{UA}, the crossover and mutation operations of the standard genetic algorithm cannot be carried out directly.
However, different groups with the same elite elements can cross each other, which is equivalent to exchanging users in different groups.
Therefore, we use the crossover operation to create new individuals from the elites as offspring.
To prevent better parent individuals from being replaced by child individuals, we engage all parents and child individuals in natural selection and choose the individuals with the highest fitness levels among them as elites.

\subsection{SUS Algorithm and Greedy Algorithm for Comparison}
Semiorthogonal user selection (SUS) \cite{Yoo2006JSAC} is a traditional efficient user scheduling algorithm in a linear beamforming system for MU-MIMO downlink transmission by ZF.
The algorithm aims to select semiorthogonal users and place them into a group to be served by ZF. simultaneously.
This algorithm has achieved significant results in MIMO scenarios, so it is widely used to compare the performance of grouping algorithms \cite{Lee2018TCOMM}.

The main idea of the greedy algorithm in \cite{Zhao2017TWC} is to traverse all users each time, simulate joining the group, select the user with the highest rate after joining the group as the user that joined the group, and repeat until all users are selected.
Different from \cite{Zhao2017TWC}, we do not have the concept of a spatial support index interval in the FD system, and to keep the number of groups fixed, we lift this restriction.

\section{Simulation and Analysis}\label{Sec6}
In this section, the numerical results are presented to verify our discussion and algorithms first.
A TDD massive MIMO system is considered in this section, where the base station is equipped with $N = 128$ antenna ULAs.
$K$ single-antenna users are uniformly located inside a fan area cell with a radius of $300\mathrm{m}$, in which the central angle is usually set as $2\pi/3$.
The path loss is given according to
\begin{subequations}
    \begin{align}
        \mathrm{PL}_{\mathrm{L}}(dB)&=-30.18+21\log_{10}(d)+20\log_{10}(f_c)+X_{\mathrm{L}}\\
        \mathrm{PL}_{\mathrm{N}}(dB)&=-34.53+34\log_{10}(d)+20\log_{10}(f_c)+X_{\mathrm{N}},
    \end{align}
\end{subequations}
where $\mathrm{PL}_{\mathrm{L}}$ for LOS paths, $\mathrm{PL}_{\mathrm{N}}$ is for NLOS paths,$X_\mathrm{L} \sim \mathcal{CN}(0,10^{0.72})$ and $X_\mathrm{L} \sim \mathcal{CN}(0,10^{1.94})$ are shadow fadings, $d$ is the distance between a user and the base station in meters and $f_c=28000\mathrm{MHz}$ \cite{Wang2021ACCESS,Fan2018TWC,Maccartney2015ACCESS}.
We set the number of NLOS paths as $L=2$.
$\rho_{\mathrm{DL}} = 50 \mathrm{dBm}$ is the downlink power constraint on the base station side, and $\sigma^2 = -104 \mathrm{dBm}$ is the noise variance on the user side.

\subsection{Angular Spacing Equalization Grouping}
\begin{figure*}[!t]
\centering
\subfigure[MRT]{\includegraphics[width=2.35in]{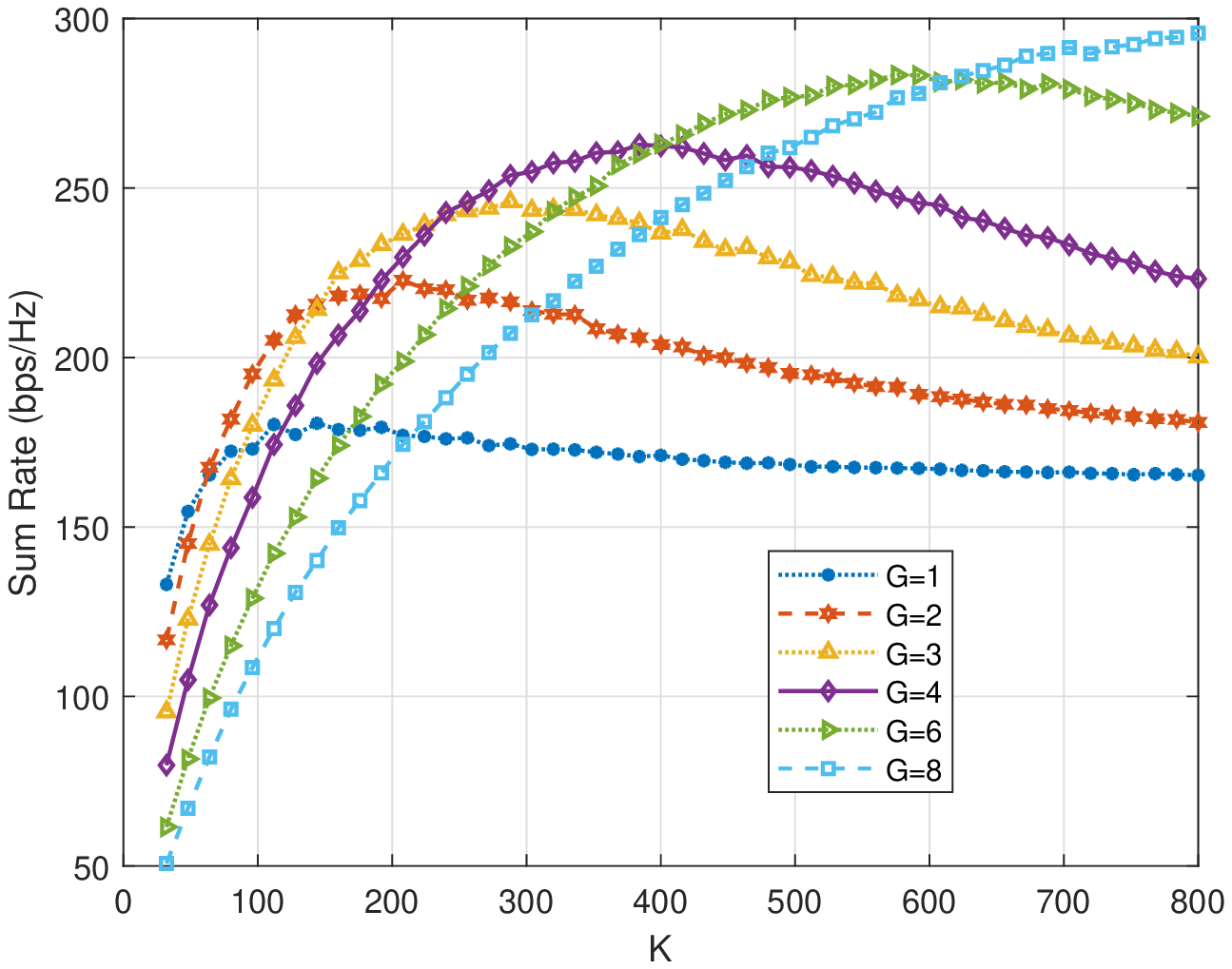}%
\label{ASEG-MRT}}
\subfigure[ZF]{\includegraphics[width=2.35in]{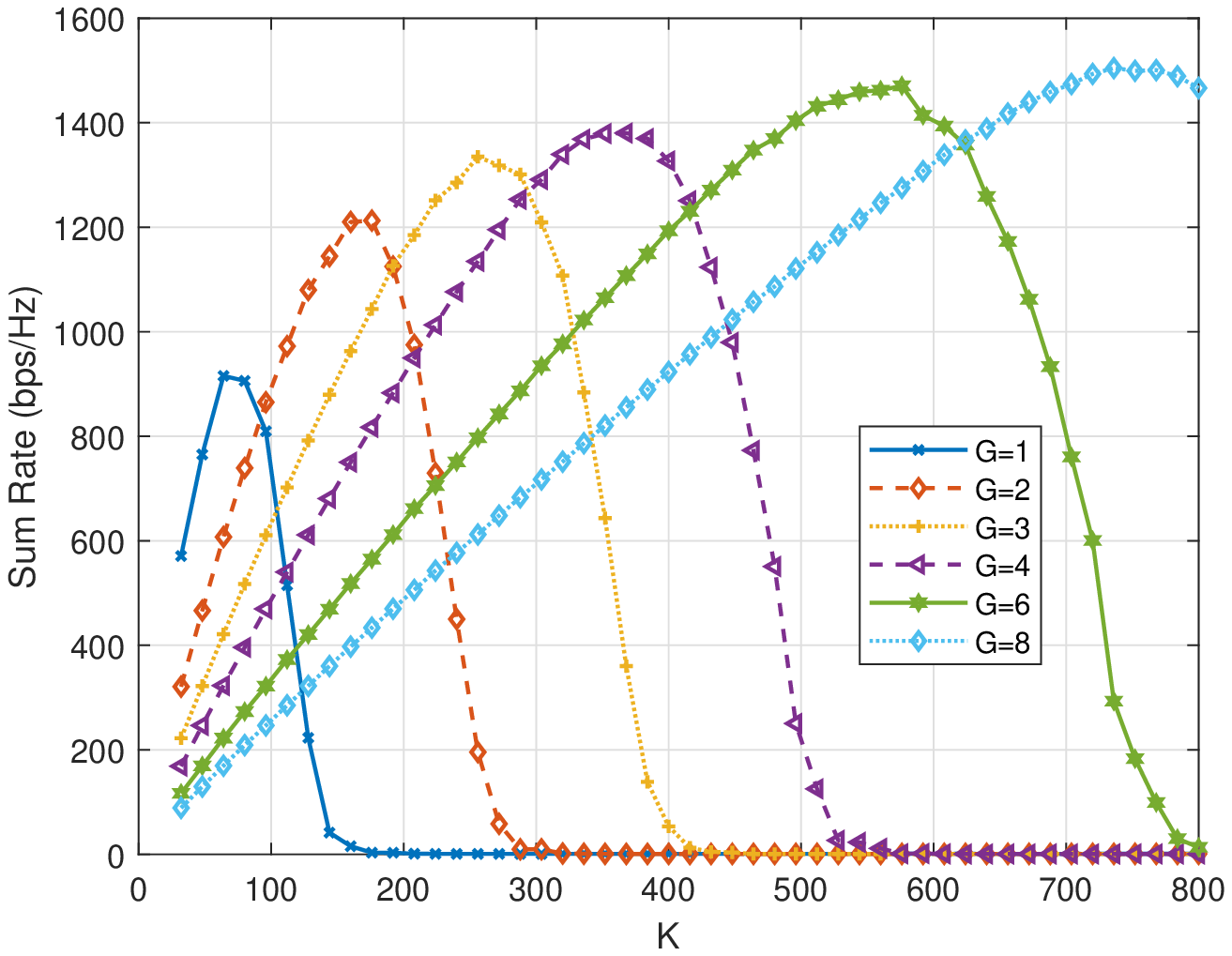}%
\label{ASEG-ZF}}
\subfigure[MMSE]{\includegraphics[width=2.35in]{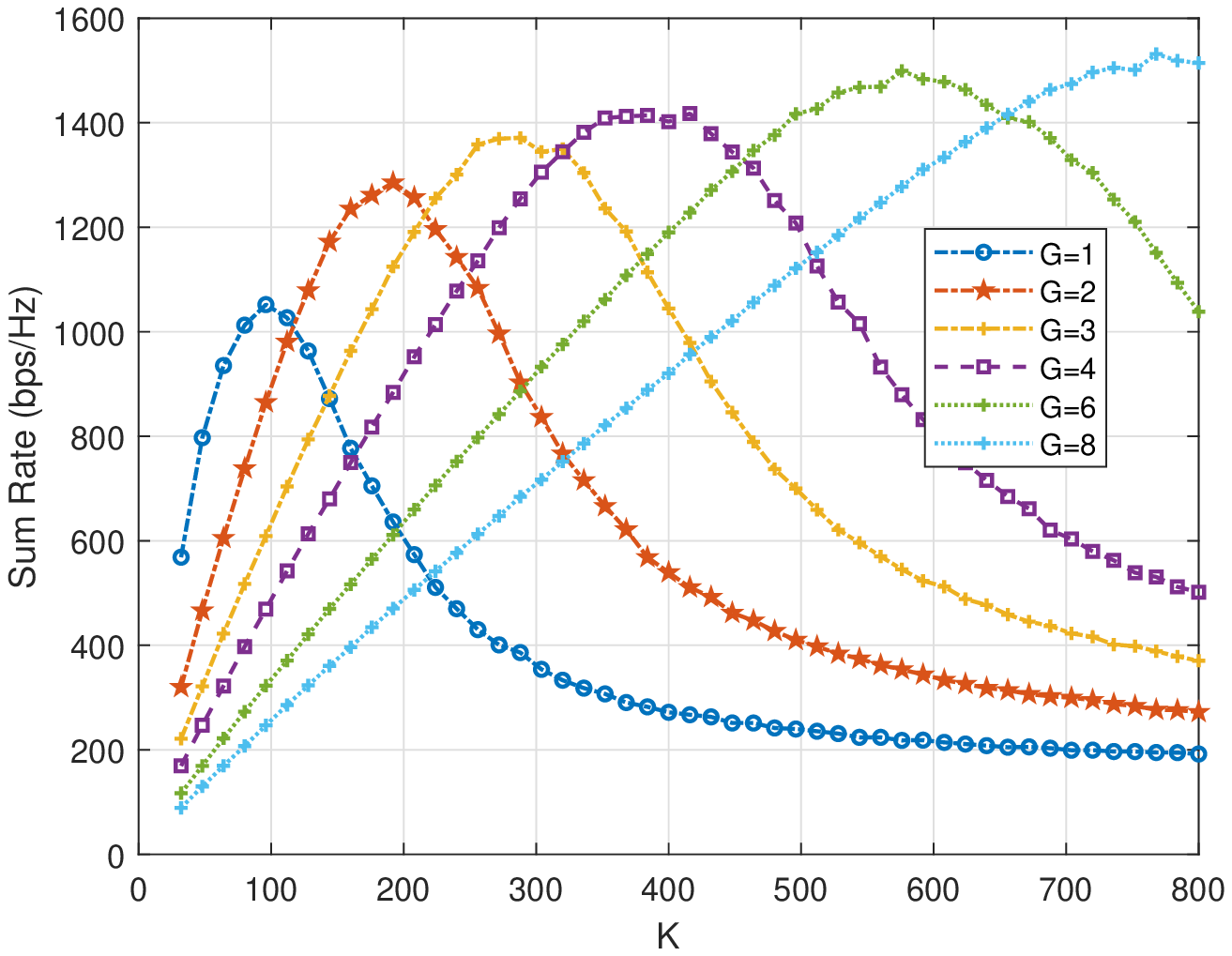}%
\label{ASEG-MMSE}}
\caption{sum-rate versus the number of groups when using ASEG.}
\label{ASEG-G}
\end{figure*}
The ASEG algorithm can improve the spectral efficiency of the system when the number of users is high and increase the maximum spectral efficiency of the system at the same time.
The case of $G=1$ in Fig. \ref{ASEG-G} is equivalent to no grouping.
The peak value of $G=1$ in Fig. \ref{ASEG-G} can be used as a reference standard to observe the improvement in the system rate obtained by a reasonable grouping.
Unlike random grouping, the ASEG grouping method sorts users according to their angles and then divides them into ordered groups.
This method divides users with larger angular spacing into the same group to make better use of the angular multiplexing gain.
The simulation results in Fig. \ref{ASEG-G} show that this method brings obvious spectral efficiency gains, indicating that an increase in angular spacing can increase the angular multiplexing gain.

Although this algorithm is proposed based on the analysis of ZF, it is still suitable for MRT and MMSE precoding and achieves good results, as shown in Fig. \ref{ASEG-G}, which shows that the algorithm has a certain universality.

\subsection{Comparison and algorithm analysis}
\begin{figure*}[!htb]
  \centering
    \subfigure[sum-rate of the system]{\includegraphics[width=0.32\textwidth]{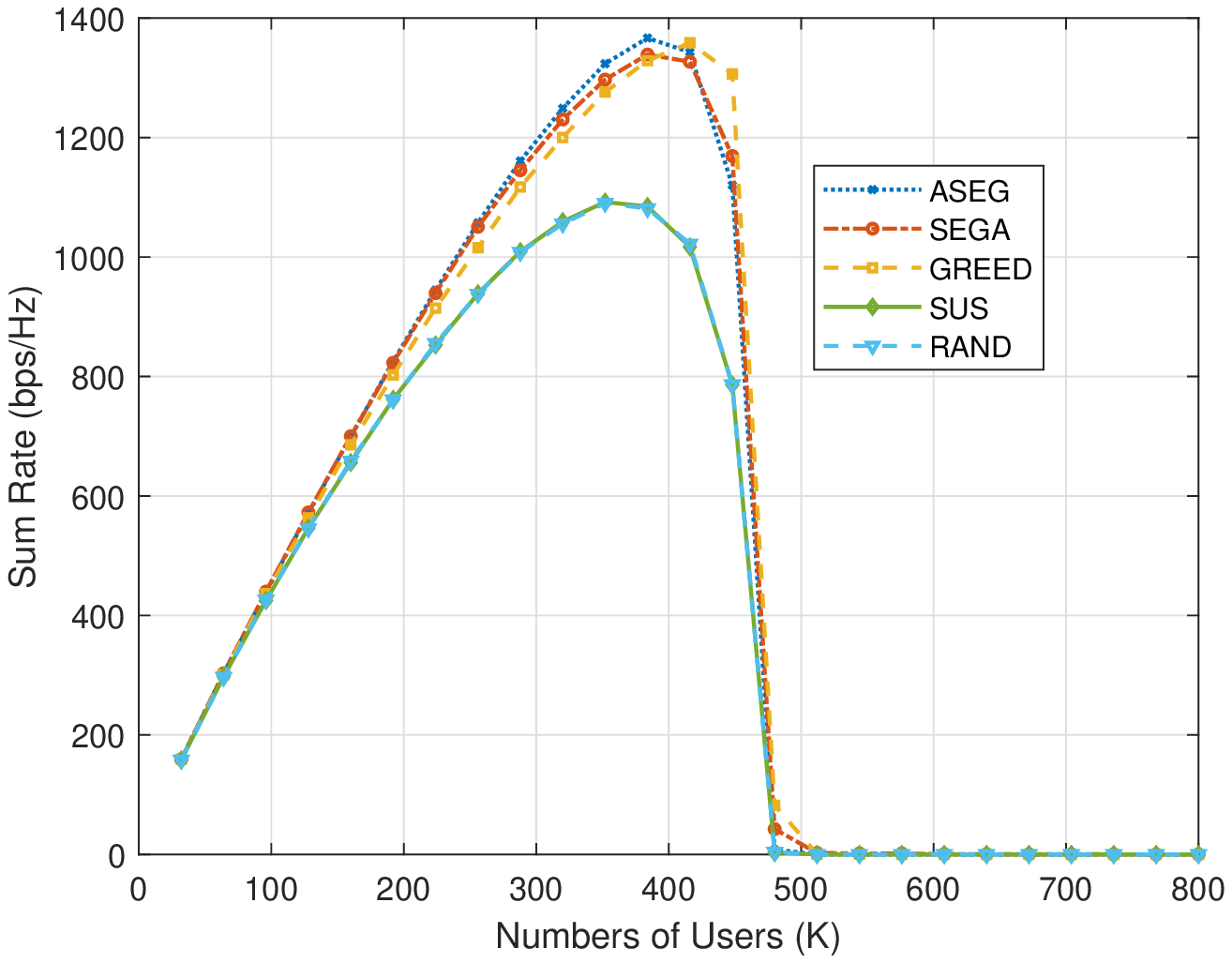}\label{COMP-ZF}}
    \subfigure[sum-rate of the system with SUS-G1]{\includegraphics[width=0.32\textwidth]{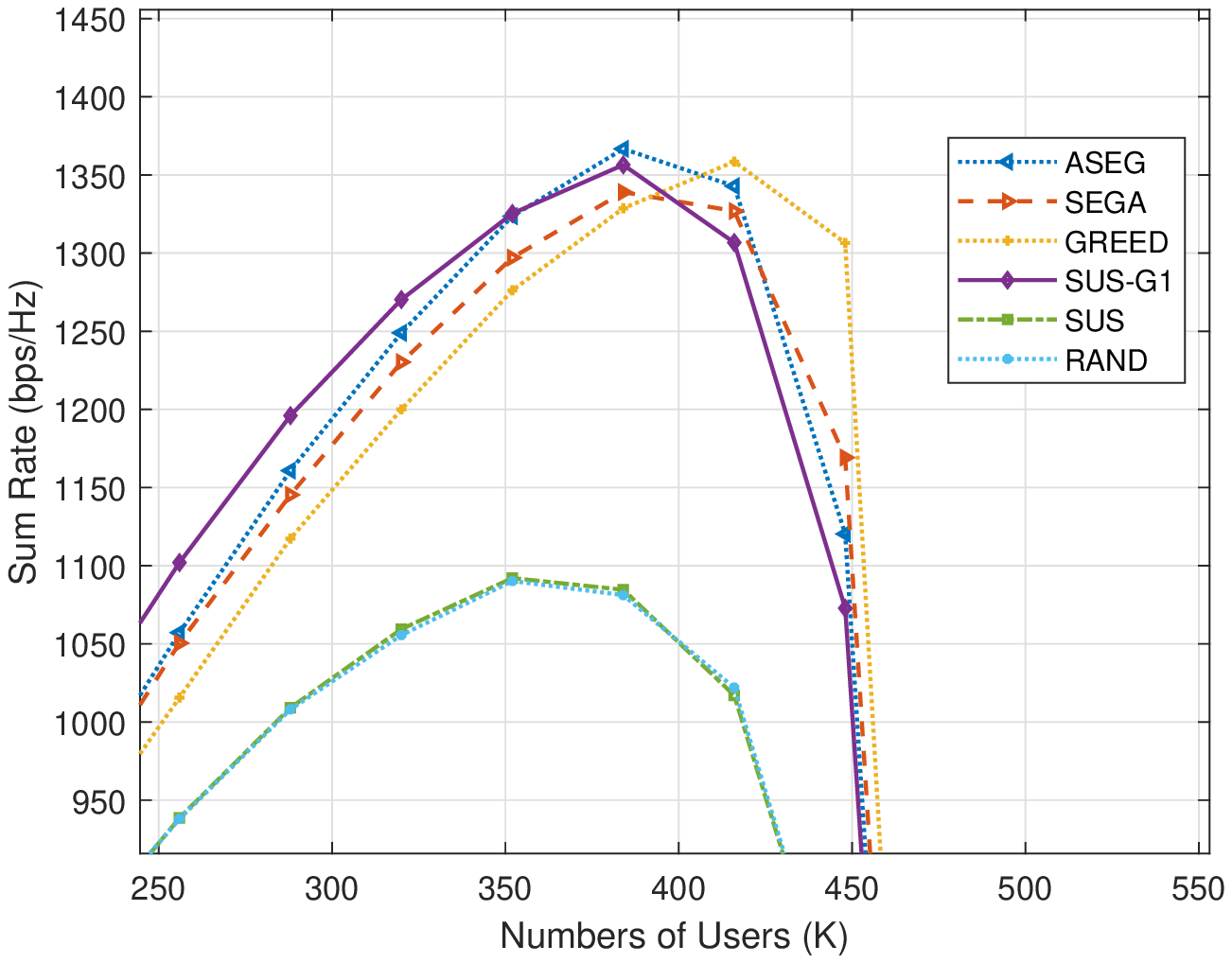}\label{COMP-ZF-D}}
    \subfigure[Time used during the simulation when ZF is used]{\includegraphics[width=0.32\textwidth]{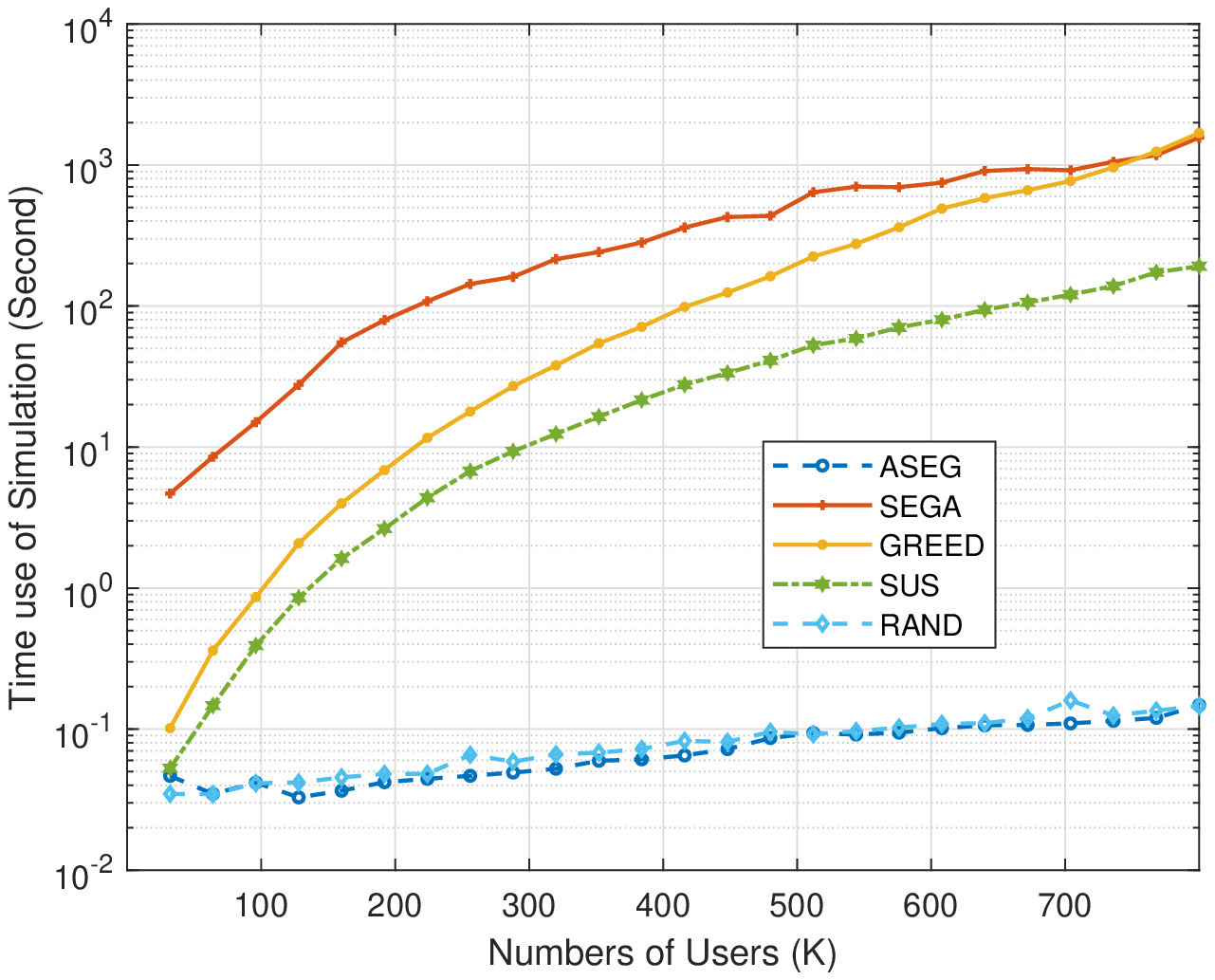}\label{COMP-TIME}}
\caption{Comparison of the rates and times of different grouping algorithms when ZF is used}
\end{figure*}
In this subsection, five different grouping algorithms are presented and their performance when ZF is used is compared when $G=4$.
Note that SUS-G1 is the first group of SUS in \cite{Yoo2006JSAC}, and GREED represents the greedy algorithm in \cite{Zhao2017TWC}.

For ZF, as shown in Fig. \ref{COMP-ZF}, ASEG has a clear advantage over other algorithms before reaching the peak.
However, it dropped more in the second half and was overtaken by the other algorithms.
This illustrates the correctness of our previous approximation behavior around its peak and that ASEG also suffers severe performance degradation when the $2/N$ assumption gradually breaks down as the number of users increases, which is to be expected.
It is worth noting that, due to the conclusion of the previous section, we can see that for the grouping problem, the most important performance indicator is the peak value because the performance drop after the peak value can be addressed by adding more groups.
The SUS algorithm still does not achieve meaningful results, and we can see that the first group of SUS does increase the user rate in Fig. \ref{COMP-ZF-D}.
Since SUS-G1 reached the highest rate in \cite{Yoo2006JSAC, Lee2018TCOMM}, here, we research other algorithms that are reasonably overlooked.

It can be seen that through the study of the actual physical performance of a communication system and the algorithm designed in a physical sense, it is possible to obtain a great comprehensive performance advantage compared to the greedy algorithm or the intelligent algorithm, which are only designed from a mathematical point of view.

Through the comparison of the above three algorithms, it can also be found that none of the three algorithms outperforms the other algorithms in all scenarios, indicating that the three algorithms only partially utilize the angular multiplexing gain under specific circumstances, which means the current research on how to exhaust the angular multiplexing gain by grouping is far from 
complete
.
We need to conduct more research on the angular multiplexing gain.

\subsection{Analysis of the algorithm complexity}
The algorithm complexity is an important indicator used to evaluate the algorithm, and it is also the key to whether the algorithm can be applied in practical scenarios.
For example, for DPC precoding, although its performance is very good, its high complexity makes it more applicable in theoretical analyses rather than in practical applications.

The most complex part of the ASEG algorithm is sorting, so its algorithm complexity is affected by the sorting algorithm used.
The research on sorting algorithms is very mature.
Generally, the algorithm complexity of traditional sorting algorithms, such as merge sort and quick sort, can easily reach $O\left( K \log_2(K) \right)$.

The algorithmic complexity of GREED is mainly composed of a greedy process, and the rate calculation $O\left( {{K}^{4}}N \right)$ far exceeds that of the ASED algorithm. The algorithmic complexity of SEGA is iterative, the elite and rate calculations require $O\left( IE({{K}^{2}}N) \right)$, where the values of $I$ and $E$ are usually on the same order of magnitude as $K$.

As a result, the ASED algorithm has an extremely low algorithm complexity, which is even lower than the complexity of multiplying two matrices, and the increase in algorithm complexity caused by user growth is not obvious.
Furthermore, the complexity of the greedy algorithm increases by the fourth power with the number of users.
Although in theory, SEGA  only doubles in complexity, to improve the performance of SEGA, the number of iterations and the amount of retention are often increased simultaneously.
The number of elites eventually leads to an increase in complexity that is close to the increase experienced by GREED as the number of users increases.

Fig. \ref{COMP-TIME} shows the average simulation time for a single simulation.
To cope with a large number of simulation tasks, we use a multicore server to perform 60 sets of Monte-Carlo simulations simultaneously with parallel computing.
Fig. \ref{COMP-TIME} shows that SEGA has the highest complexity, followed by GREED and SUS.
The complexity of ASEG is not significantly different from RAND and is much lower than that of the other algorithms.
The complexity of these algorithms increases rapidly with the increase in users and is three to four orders of magnitude higher than the complexity of ASEG when the number of users is large.
Fig. \ref{COMP-TIME} also shows the extremely low algorithmic complexity of the ASEG algorithm in comparison to that of the other algorithms. 
The algorithmic complexity of ASEG is not meaningfully different from that of RAND, which means that the increased complexity of the ASEG algorithm is even partially masked by other simulation contents, such as the precoding calculation and rate calculation.

In general, ASEG has extremely low and even negligible algorithm complexity while maintaining excellent performance; hence, the ASEG algorithm has potential practical application value.

\section{Conclusion} \label{Sec7}
In this paper, we investigated the angular multiplexing gain and proposed a grouping scheme for ADMA mixed with a TDMA system.
Through the study of the physical beamforming effect of precoding in mmWave massive MIMO systems, we introduced the relationship between the angular multiplexing gain and the angle range distribution of users.
By the above theory of the angular multiplexing gain, we transformed the sum-rate maximization problem for ADMA systems into the problem of equalizing user angular spacing, proposed the ASEG algorithm, and proved that it is optimal for the above approximation problems.
The simulation results show that this algorithm has very low complexity and high performance.

\bibliographystyle{IEEEtran}

\bibliography{ref}

\end{document}